\documentclass[aps,showpacs,twocolumn]{revtex4}
\usepackage{epsfig}
\usepackage{graphics}
\usepackage{amssymb}
\usepackage{amsmath}

\newcommand{\intsum}[1]{\sum_{#1} \! \! \! \! \! \! \! \! \! \int }
\begin{document}
\title{Attosecond delays in laser-assisted photodetachment \\ from closed-shell negative ions}
\author{Eva Lindroth and Jan Marcus Dahlstr{\"o}m}
\affiliation{Department of Physics, Stockholm University, AlbaNova University Center, SE-106 91 Stockholm, Sweden}

\begin{abstract}
We study laser-assisted photodetachment time delays by attosecond pulse trains from the closed-shell negative ions F$^-$ and Cl$^-$. We investigate the separability of the delay into two contributions: (i) the Wigner-like delay associated with one-photon ionization by the attosecond pulse train and (ii) the delay associated with exchange of an additional laser photon in the presence of the potential of the remaining target. Based on the asymptotic form of the wave packet, the latter term is expected to be negligible because the ion is neutralized leading to a vanishing laser--ion interaction with increasing electron--atom separation. While this asymptotic behavior is verified at high photoelectron energies, we also quantify sharp deviations at low photoelectron energies. Further, these low-energy delays are clearly different for the two studied anions indicating a breakdown of the universality of laser--ion induced delays. The fact that the short-range potential can induce a delay of as much as 50 as can have implications for the interpretation of delay measurements also in other systems that lack long-range potential. 
\end{abstract}

\maketitle

% pacs

\section{Introduction}
There is a renewed interest in atomic and molecular photoionization and quantum control driven by the rapid development of coherent light sources in the extreme ultraviolet (XUV) and soft x-ray range \cite{ChiniNaturePhotonics2014}. Attosecond XUV pulses phase-locked to laser fields have been used to study bound electron dynamics in atoms \cite{GouliemakisNature2010} and autoionization processes \cite{KaldunNature2016} in real time. Temporal characterization of attosecond pulses, which requires determination of the spectral phase of the XUV field, can be performed by laser-assisted photoionization \cite{PaulScience2001,ItataniPRL2002}. More recently, this type of experiments have been used for spectral phase determination of photoelectrons from autoionizing states \cite{KoturNatCom2016,GrusonScience2016} and for the measurement of relative time delays in laser-assisted photoionization from different initial states in solids~\cite{CavalieriNature2007} and atoms \cite{SchultzeScience2010,KlunderPRL2011,GuenotPRA2012}. Relative time delays have also been measured between different atomic species~\cite{PalatchiJPB2014,GuenotJPB2014,Sabbar:2015}, between single and double ionization \cite{ManssonNP2014}, between different angles of photoemission \cite{HeuserPRA2016} and between the ion ground state and shake-up states \cite{Ossiander2017}. In the case of atomic photoionization it has been established that the measured atomic delay $\tau_A$ can be separated into a Wigner-like delay \cite{eisenbud:thesis,wigner:timedelay,smith:timedelay}, $\tau_W$, here associated with the one-photon XUV ionization process, and a contribution from the interaction with the laser and long-range ionic field, $\tau_{CC}$, called the continuum-continuum delay (or Coulomb-laser coupling delay) \cite{ZhangPRA2010,DahlstromJPB2012,PazourekFD2013}. The $\tau_{CC}$ term is ``universal'' in the sense that it depends only on the kinetic energy of the photoelectron, the photon energy of the laser field and the charge of the remaining ion ($Z=1$ for neutral targets). The validity of the relation, $\tau_A\approx\tau_W+\tau_{CC}$, has been demonstrated theoretically by ab initio calculations from the ground states in hydrogen \cite{NagelePRA2011,DahlstromCP2013} and helium \cite{PazourekPRL2012} and from inner or outer valence orbitals of noble gas atoms using diagrammatic perturbation theory \cite{DahlstromPRA2012,DahlstromJPB2014,DahlstromJPB2014corr}. Depending on the target an additional correction term due to dipole--laser coupling will arise in atomic systems with (nearly) degenerate states or permanent dipoles \cite{PazourekFD2013}. In the language of perturbation theory similar effects are expressed by the reversed time-order of photon interactions. While it has been understood for a long time that the simple additive relation is not valid close to atomic resonances~\cite{JimenezPRL2014,Sabbar:2015,Kotur:2016}, it has now been experimentally shown that the simple relation breaks down also in angular-resolved measurements from the isotropic ground state of helium \cite{HeuserPRA2016}. The latter effect is especially strong for photoelectron emission close to the nodes of the photoelectron angular distributions where the probability of electron emission is low. Other recent work on photoionization time delays include the measurement and interpretation of attosecond laser-assisted delays from anisotropic molecular targets \cite{HockettJPB2016,HuppertPRL2016,BaykushevaCP2017,Baykusheva2017comment}. 

Emission of an electron from a neutral atom (or positive ion) by absorption of a photon is called photoionization, while neutralization of a negative ion by emission of an electron is called photodetachment. 
The analytical formula for $\tau_{CC}$, given in Eq.~(100) of Ref.~\cite{DahlstromJPB2012}, depends on the long-range Coulomb potential and it predicts that $\tau_{CC}=0$ for negative ions. In more detail, this result was derived using the asymptotic form of continuum wavefunctions, within Wentzel-Kramers-Brillouin (WKB) theory with angular momentum $\ell$, given by   
\begin{align}
u_{k}(r)=N_k(r)\exp\{i[kr+\Phi_k(r)]\},
\label{uWKB}
\end{align}
where the correct asymptotic phase shift of the photoelectron was imposed, $\Phi_k(r)=Z \ln(2kr)/(ka_0)+\eta_\ell(k)-\pi\ell/2$, given a short-range phase shift, $\eta_\ell(k)$, and where $N_k(r)$ is the WKB amplitude that depends on the local momentum \cite{DahlstromCP2013}. At first glance, the success of this approximation is surprising because it implies an inaccurate wavefunction description at the core where the potential is the strongest, but it was found that the CC-delay could be explained in hydrogen by dipole transitions between these asymptotic states with different logarithmic phase shifts. If instead the potential is short ranged, $Z=0$, then there will be no logarithmic phase-shifts in Eq.~(\ref{uWKB}) and it is not clear how well the asymptotic approximation will work. However, the prediction that the CC-delay is small is supported by time-dependent calculations using short-range model potentials (of Yukawa type) for photoelectrons with high energy \cite{NagelePRA2011}. A similar situation may occur for ionization from surfaces, or surface adsorbates if the charge left behind in the remaining system is swiftly and efficiently screened. 

\begin{figure}
\includegraphics[width=0.490\textwidth]{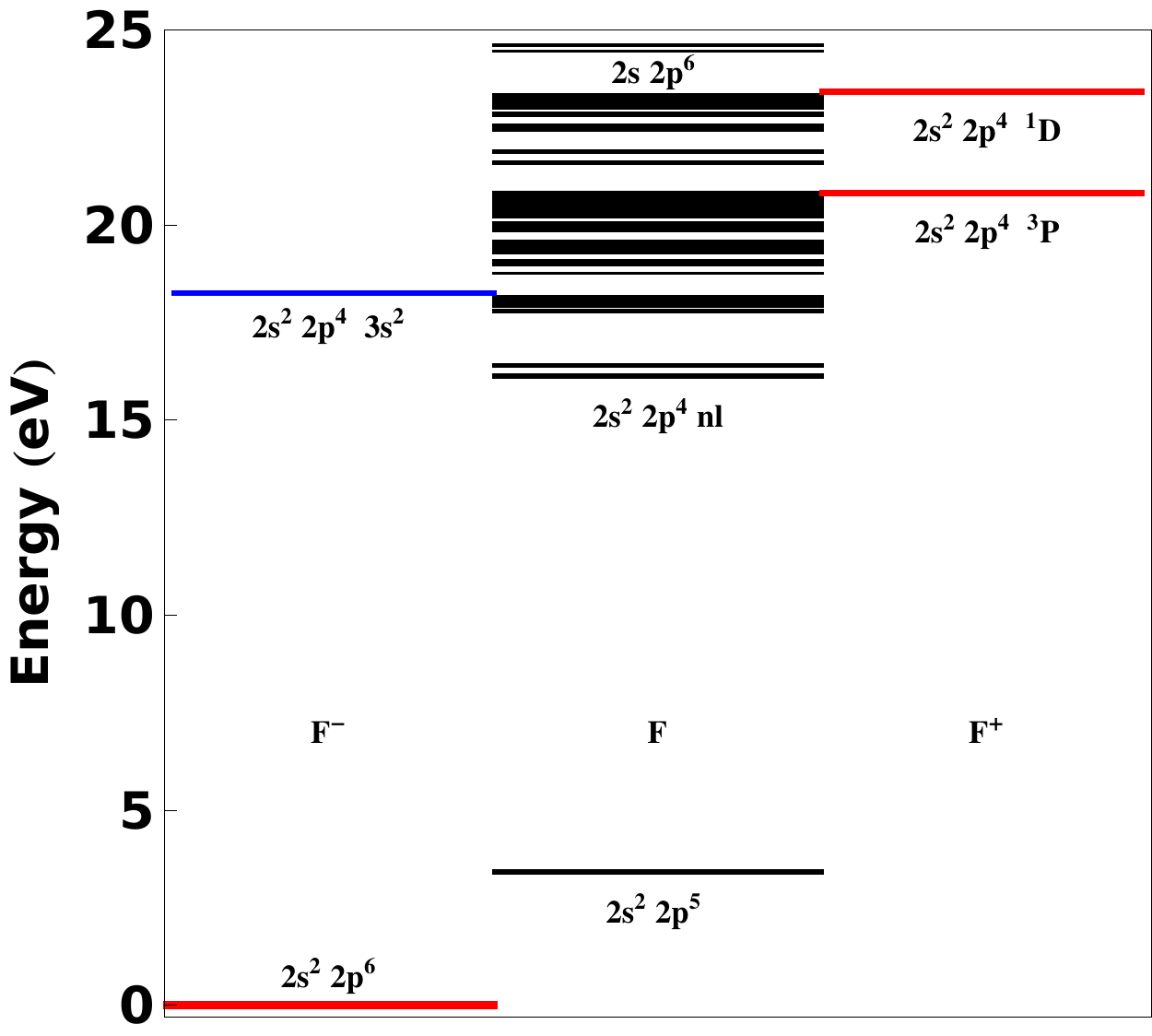}
\includegraphics[width=0.490\textwidth]{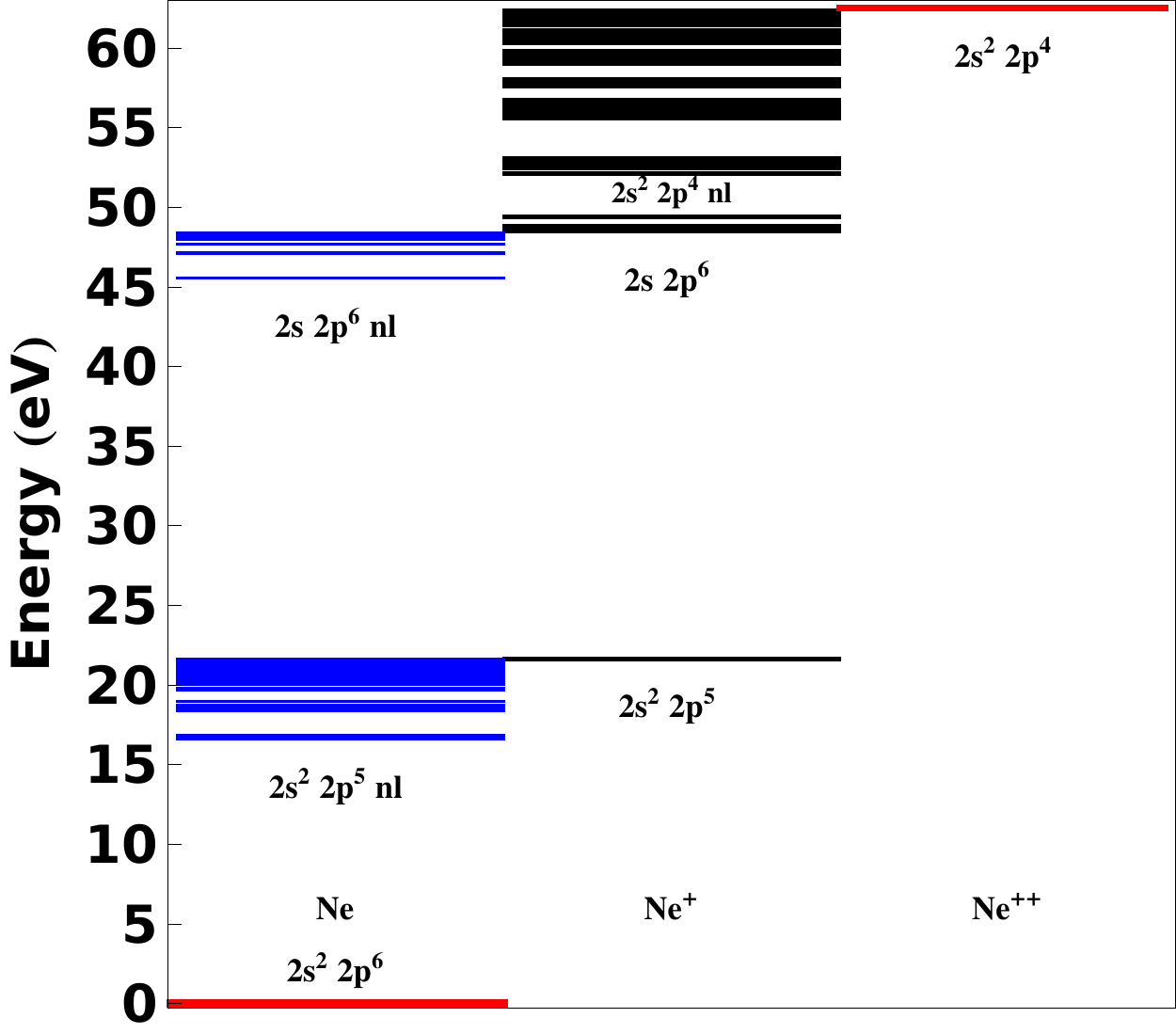}
\caption{The energy levels~\cite{NIST:database} in the Ne-like, F-like and  O-like systems with $Z = 9$ (upper figure) and $Z= 10$ (lower figure).}
\label{Fig:ne}
\end{figure}

\begin{figure}
\includegraphics[width=0.490\textwidth]{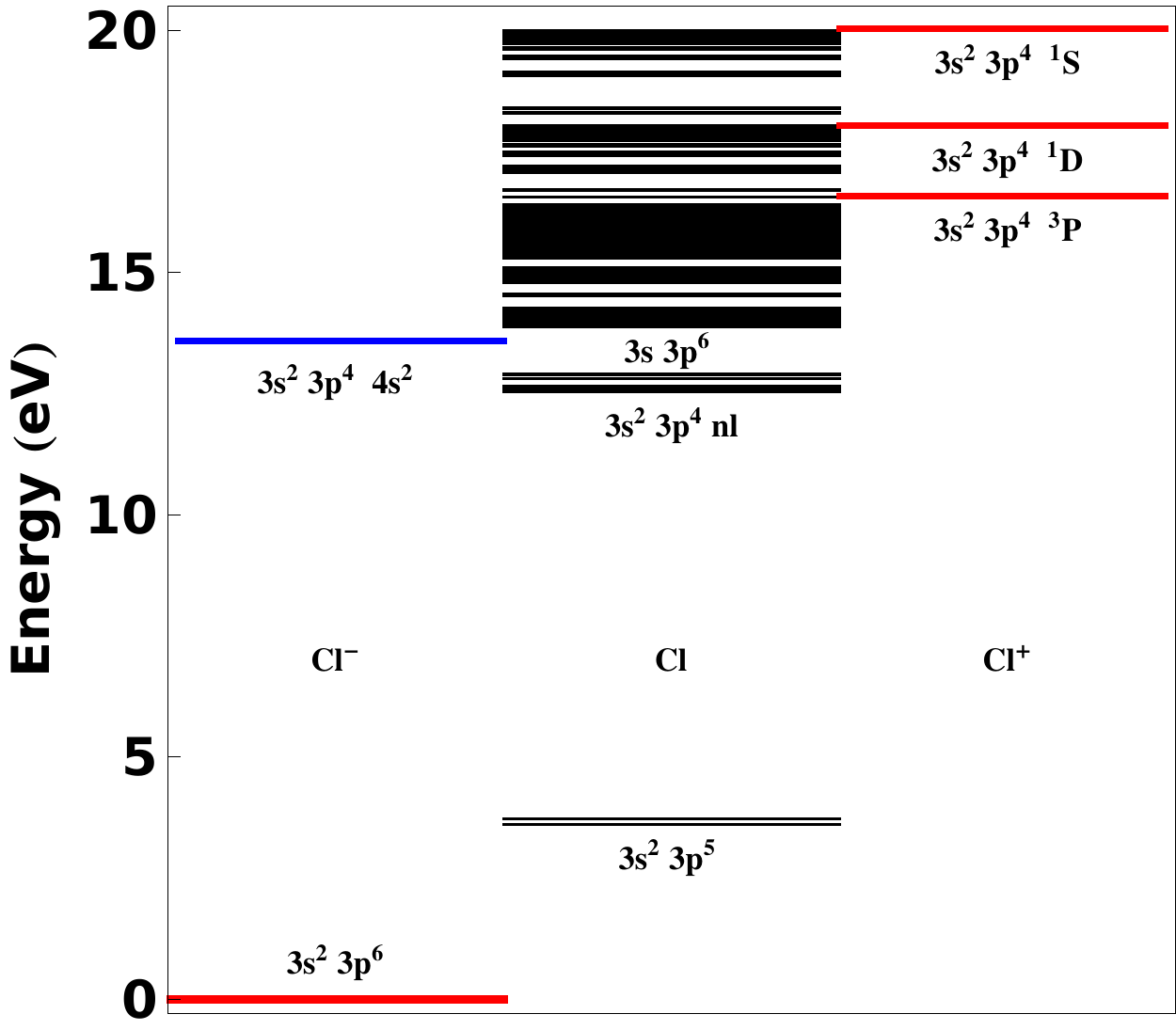}
\includegraphics[width=0.490\textwidth]{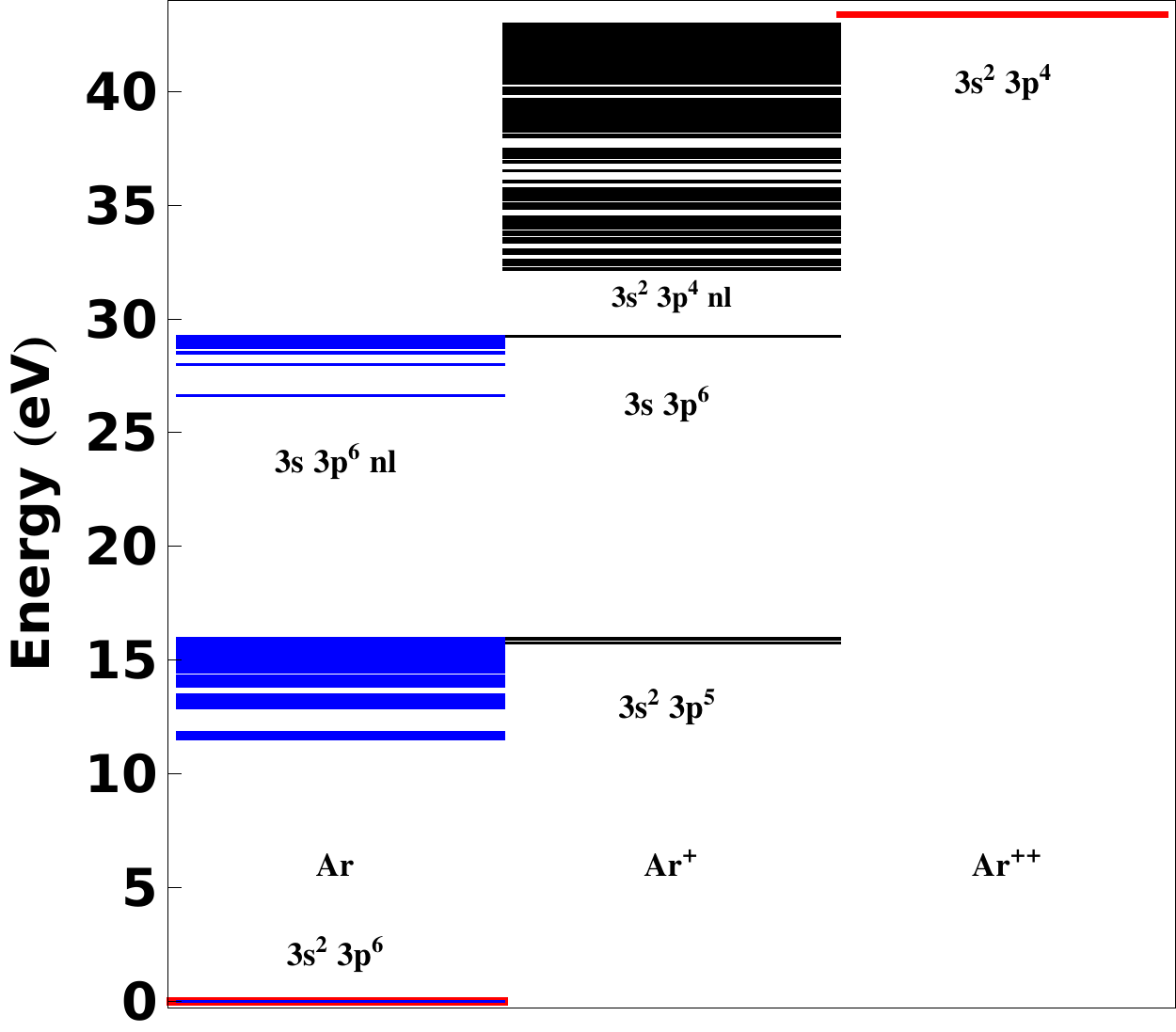}
\caption{The energy levels~\cite{NIST:database} in the Ar-like, Cl-like and  S-like systems with $Z = 17$ (upper figure) and $Z=18$ (lower figure).}
\label{Fig:ar}
\end{figure}

In this paper, we present work on laser-assisted photodetachment of negative halogen ions with the aim to answer the question if it is possible to measure directly the photoelectron Wigner delay $\tau_W$ from targets that lack long-range interaction. We have chosen to study the negative halogen ions of fluorine and chlorine for two reasons (i) they both have large affinities and (ii) they are isoelectronic to the much studied rare gases atoms neon and argon. In spite of having identical ground state configurations as neon (argon), the negative fluorine (chlorine) ion exhibit important differences due to the missing long-range Coulomb potential. While neutral atoms supports an infinite number of Rydberg states converging to the ionization threshold, these negative ions have no bound excited states and only a few identified autodetaching resonances \cite{Berzinsh:1995,Torkild:review}. 
In F$^-$ and Cl$^-$ only one autodetaching resonance has been found corresponding to the $\ldots 2p^4 3s^2$  and $\ldots 3p^4 4s^2$-configurations, respectively \cite{Torkild:review}. At the second ionization threshold in neon and argon the remaining ion is left with a single $s$-hole with dominating configurations $\ldots 2s 2p^6$ and $\ldots 3s 3p^6$, respectively. Doubly excited states in neon and argon, leading to  $\ldots 2s^2 2p^4 n\ell$ and $\ldots 3s^2 3p^4 n\ell$ ion configurations, are found at higher energies. In fluorine the order is reversed and nearly all of the states dominated by the $\ldots  2s^2 2p^4 n\ell $-configurations are below $\ldots 2s 2p^6$-configuration, which further is autoionizing since it is above the ground state of F$^+$. 
The situation is less drastically changed  in Cl$^-$  where only some of the $\ldots 3s^2 3p^4 n\ell$-states are below the $\ldots 3s3p^6$-threshold, which  is stable against further decay to Cl$^+$. The energy levels of the discussed systems are shown in Figs.~(\ref{Fig:ne}-\ref{Fig:ar}) \cite{NIST:database}. 
Here we are not primarily interested  in the individual properties of these particular negative ions, but more in the overall size and energy dependence of the continuum-continuum delay in a system without long-range potential. The focus is on photoelectron emission from the outermost shell where the Random Phase approximation with Exchange (RPAE) is known to produce good results for photoionization cross sections of rare gas atoms \cite{Amusia1990} and negative halogen ions \cite{RodojevicPRA1987}. It is further well tested that, with the exception of the immediate threshold regions, only minor contributions come from inter-shell couplings in length gauge. 
The paper is organized as follows. 
In Sec.~\ref{theory} we briefly review the theory for atomic delays and in 
   Sec.~\ref{method} we review our numerical method. 
Finally, our results and conclusions are presented in Sec.~\ref{results} and \ref{sec:conclusions}, respectively.

\section{Theory}
\label{theory}
Here we will briefly discuss the calculation of delays in laser-assisted photoionization and photodetachment. A more detailed account can be found in Ref.~\cite{DahlstromJPB2014}. 
We consider first an $N$-electron atom or negative ion that absorbs one photon and subsequently ejects a photoelectron. 
%that asymptotically has angular momentum $\ell$ and energy $\hbar^2\kappa^2/2m$. 
The radial photoelectron wavefunction will asymptotically be described by an outgoing phase-shifted Coulomb wave,

\begin{align}
\label{onephoton2}
\nonumber \\
u_{\rho,\ell}^{[1]}
%_{k,\ell} 
\left( r \right) \approx - \pi M^{(1)} \left(k_{\ell}, \Omega, a \right) 
E_{\Omega}  
\sqrt{\frac{2m}{\pi k \hbar^2}} \nonumber \\
 \times e^{i \left( k r + \frac{Z}{k a_0} \ln 2k r- \ell \frac{\pi}{2} + \sigma_{\ell,Z}\left( k \right) + \delta_{\ell}\left( k \right)
 \right) },
\end{align}
where $\sigma_{Z,k,\ell}$ is the Coulomb phase with charge $Z$ for a photoelectron with wave number $k$ and angular momentum quantum number $\ell$ given by   
\begin{align}
\sigma_{\ell, Z} \left( k \right)= \arg \left[ \Gamma  \left(  \ell + 1  -i\frac{Z}{k a_0} \right) \right].
\end{align}
Negative ions have $Z=0$ which implies that Coulomb phase is zero, while neutral atoms have $Z=1$ which equals the asymptotic phase shifts of hydrogen.  
Additional phase-shifts include short-range corrections due to the static atomic potential $\delta_{\ell}\left( k \right)$ and the phase-shift of the XUV  field and correlation effects in $M^{(1)}$. The non-correlated transition matrix element in length gauge with a linearly polarized field, $E_{\Omega}$ in Eq.~(\ref{onephoton2}), is given by   
\begin{align}
\label{M1}
M^{(1)} \left( k_\ell, \Omega, a \right) =  \langle k_\ell \mid  e z \mid a\rangle,
\end{align}  
where the dipole transition $\langle k_\ell \mid  e z \mid a\rangle$ can be chosen to be real describing a transition from the initial occupied state $a$ with $(n_a,\ell_a,m_a)$ to the final state $(k,\ell,m_a)$. In the following we will denote the full perturbed wavefunction associated with absorption of one photon with angular frequency $\Omega$ and a hole in $a$ by $\left|\rho_{\Omega,a}\right>$ including both radial, angular and spin parts implicitly.  

%We can call the sum of these two phases $\eta$
%\begin{align}
%\eta_{\ell, Z}  \left( \kappa \right) = \sigma_{\ell,Z}\left( \kappa \right)  + \delta_{\ell} \left( \kappa \right),
%\end{align}
%together with the angular momentum phase in Eq.~(\ref{onephoton2})  $\eta$ gives the total asymptotic  phase shift, $\phi^{(1)}$, relative the $r$-dependent wave function, of the photoelectron, i.e.
%\begin{align}
%\label{phi1}
%\phi^{(1)}=
%-\ell \frac{\pi}{2} + \eta_{ \ell, Z} \left( \kappa \right),
%\end{align}
%where the number in the superscript refer to the number of absorbed photon. The energy derivative of this phase shift gives the delay of the electron travelling  through the atomic potential as compared to a free particle\cite{eisenbud:thesis,wigner:timedelay,smith:timedelay}:
%\begin{align}
%\label{twigner}
%\tau_{W,\ell} = \hbar  \frac{d\eta_{ \ell, Z}  \left( \kappa \right) }{dE} 
%\end{align}
%$\tau_{W,\ell}$ we label as the Wigner delay in angular momentum channel $\ell$. The angular resoved case will be discussed below.

We will consider measurements that employ Reconstruction of Attosecond  Beating By Interference of Two-photon Transitions (RABBITT) \cite{PaulScience2001}, where an XUV comb of odd-order harmonics of a fundamental laser field, $\Omega=(2n+1)\omega$, is combined with a synchronized, weak $\omega$ laser field. In the RABBITT setting the one-photon ionization process is assisted by an IR photon that is either absorbed or emitted. This gives rise to quantum beating of sidebands in the photoelectron spectrum at energies corresponding to the absorption of an even number of IR photons. 
%Since the absorption of  neighboring harmonics combined with either absorption or emission of  an IR photon constitute different pathways between the same initial and final quantum state, the side-band signal will oscillate as a function of  the  delay between the two fields.  
The outgoing radial wave function after interaction with two photons in resonance free region will asymptotically have the form 
\begin{align}
\label{twophoton}
u_{\rho,\ell}^{[2]}
%_{k,\ell} 
\left( r \right) 
\approx  
 -\pi
M^{(2)}(k_\ell, \pm\omega, \Omega, a)
E_{\pm\omega} E_{\Omega}
\sqrt{\frac{2m}{\pi k \hbar^2}} \nonumber \\
e^{i \left( k r + \frac{Z}{k a_0} \ln 2 kr - {\ell} \frac{\pi}{2} + \sigma_{{\ell},Z}\left( k \right) + \delta_{{\ell}}\left( k \right)
 \right) }, 
\end{align}
where the difference compared to the one-photon case lies in the presence of the two-photon transition element $M^{(2)}$, 
and in the additional field amplitude for absorption or emission of an IR photon, $E_{\pm\omega}$. 
The two-photon matrix element connects the initial state $a$ to the continuum state $k_\ell$ through all dipole-allowed intermediate states $p$, 
%As before the photo-electron is characterized by $k'$ and $\ell'$, and  $\alpha$ denotes the state of the target ion:
\begin{align}
\label{M2}
M^{(2)}(k_\ell, \omega, \Omega, a ) =
\nonumber \\
%\times
\lim_{\varepsilon \rightarrow 0^+}
\intsum{p} \frac{
\langle k_\ell \mid  e z \mid p  \rangle \langle p  \mid    e z  \mid a \rangle}{\epsilon_a + \hbar\Omega -\epsilon_p + i \varepsilon},
\end{align}
with final photoelectron energy $\epsilon_k=\epsilon_a+\Omega\pm\omega$. An important difference compared to one-photon absorption is that the two-photon matrix element is intrinsically complex even if correlation effects are excluded for the case of above-threshold ionization, $\hbar\Omega>|\epsilon_a|$. 
%and thus the phase shift after two-photons differs from that after one photon in Eq.~(\ref{phi1}) and has an additional contribution from the matrix element $M^{(2)}$:
%\begin{align}
%\phi^{(2)}_{\ell'} \left( k' \right)  = \arg \left[ M^{(2)}\right] - {\ell'}\frac{\pi}{2} +\eta_{{\ell'},Z}\left( k \right) 
%\end{align}

The atomic contribution to the quantum beating of the side band at energy $2n \hbar \omega$  in a RABBITT experiment is the phase difference between the  quantum paths where the XUV harmonic $\hbar \Omega_> = \left(2n+1\right)\hbar \omega $ is absorbed and an  IR-photon is emitted and that where both an XUV harmonic, now of energy $\hbar \Omega_< = \left(2n-1\right)\hbar \omega$, and an IR-photon is absorbed. In addition one must also consider the reversed time-order processes, where the IR photons are exchanged before absorption of any XUV photon. While it is widely known that the latter time-order plays a minor role for noble gas atoms, one can expect that the effect is more significant in laser-assisted photodetachment because affinities of negative ions are significantly smaller than  ionization thresholds of noble atoms.

Following the usual RABBITT formalism \cite{DahlstromJPB2014}, we construct the phase differences for the two quantum paths leading to emission of a photoelectron along the common polarization axis of the fields,
\begin{align}
\label{deltaphi}
\Delta \phi_{\theta=0}^{(2)}   =
 \arg 
\left[ 
\left( \sum_{\ell'} 
M^{(2e)}
e^{i 
\left( - \ell' \frac{\pi}{2} + \eta_{Z,k,\ell'} 
\right) }
 Y_{\ell',0}\left( \theta,0 \right) \right) 
\right.
\nonumber \\
\left.
  \times 
\left( \sum_{\ell''}
M^{(2a)}
e^{i 
\left( - \ell'' \frac{\pi}{2} + \eta_{Z,k,\ell''}
\right) } Y_{\ell'',0}\left( \theta,0 \right) 
\right)^*
\right] 
\end{align}
where we use the following short-hand notation
\begin{align}
M^{(2e)} &=  M^{(2)}(k_{\ell'}, -\omega, \Omega_>, a ) \nonumber \\
M^{(2a)} &=  M^{(2)}(k_{\ell''}, \omega, \Omega_<, a ), \nonumber \\
\eta_{Z,k,\ell} &= \sigma_{Z,k,\ell} + \delta_{k,\ell}, \nonumber
\end{align}
so that the atomic delay can be calculated for sideband $2n$ as 
\begin{align}
\label{atomicdelay}
\tau_a= \frac{\Delta \phi_{\theta=0}^{(2)}}{2 \omega}.
\end{align}
While we could in principle compute the atomic delay in any angle \cite{DahlstromJPB2014,DahlstromJPB2014corr}, we choose here to focus on photoelectron emission along the $z$-axis where the separation of the Wigner delay and the CC-delay is known to be accurate for noble gas atoms.  
Similarly the one-photon dipole phases in the $\theta=0$ direction  
\begin{align}
\label{deltaphiWigner}
\phi_{\theta=0}^{(1,\kappa_>)} =
 \arg 
\left( \sum_{\ell} 
M^{(1)}_{\Omega_>}
e^{i 
\left( - \ell \frac{\pi}{2} + \eta_{Z,\kappa_>,\ell} 
\right) }
 Y_{\ell,0}\left( \theta,0 \right) \right) 
\nonumber \\
\phi_{\theta=0}^{(1,\kappa_<)}   =  
\arg   
\left( \sum_{\ell}
M^{(1)}_{\Omega_<}
e^{i 
\left( - \ell \frac{\pi}{2} + \eta_{Z,\kappa_<,\ell}
\right) } Y_{\ell,0}\left( \theta,0 \right)
\right),
\end{align}
can be used to compute the Wigner-like delay at sideband $2n$ as
\begin{align}
\label{wignerdelay}
\tau_W= \frac{\phi_{\theta=0}^{(1,\kappa_>)}-
\phi_{\theta=0}^{(1,\kappa_<)}}{2 \omega}. 
\end{align}
We will refer to the quantity $\tau_a-\tau_W$ as the delay difference introduced by the second photon. 


\section{Method} 
\label{method}
The numerical implementation has been discussed in more detail 
in Ref.~\cite{DahlstromJPB2014} and will only be briefly reviewed here. 

\begin{figure*}
\includegraphics[width=0.98\textwidth]{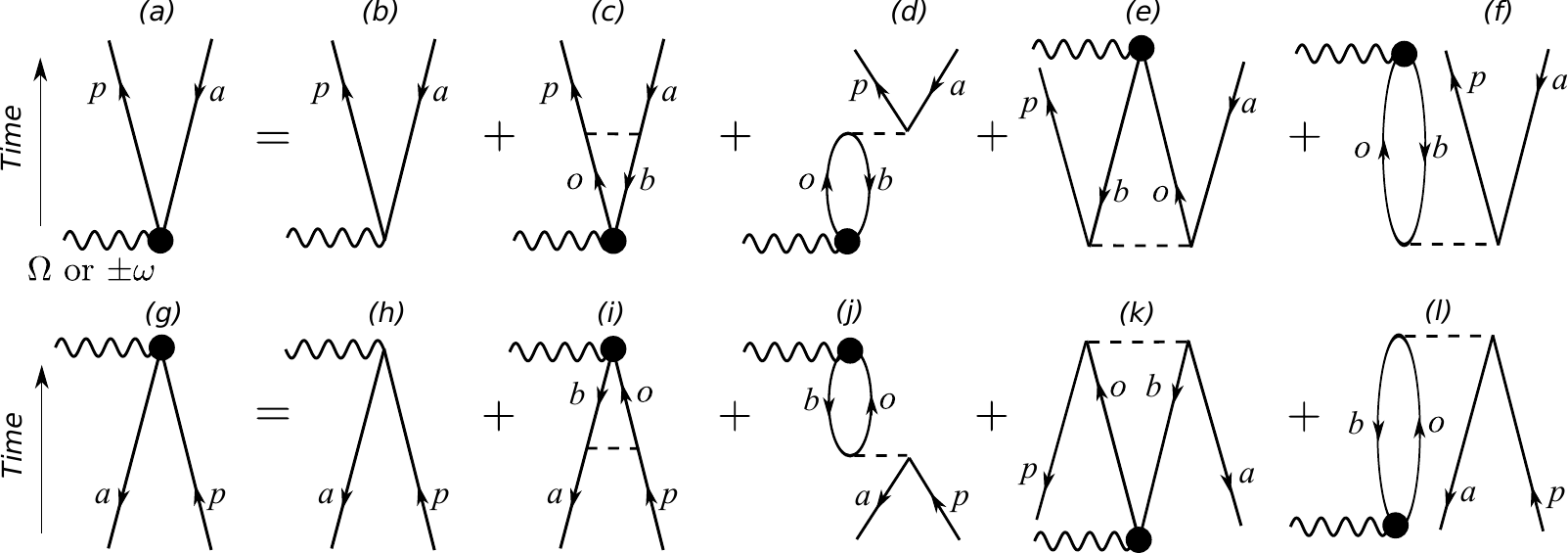}
\caption{RPAE for the many-body screening  of the photon interaction. (a) and (g) are forward and backward propagation, respectively, where the sphere indicates the correlated interaction to infinite order.
\label{fig:rpae}}
\end{figure*}

\subsection{Treatment of Many-Body Effects}
To describe %the radial wave function for  
the one-photon perturbed wave function $\rho_{\Omega,a}$  
%cf.~Eq.~(\ref{onephoton2}), 
we use the Random-Phase Approximation with Exchange (RPAE)~\cite{Amusia1990} in a basis set obtained through diagonalization of an effective one-particle Hamiltonian in a radial primitive basis of B-splines~\cite{deboor} on a complex scaled radial grid in a spherical box. We use external complex scaling (ECS)  
\begin{align}
\label{complexrotation}
r \rightarrow
\Bigg \{
\begin{array}{lr}
r, & 0 < r  < R_C\\
Rc + \left(r-R_C\right) e^{i \theta}, & r > R_C. 
\end{array}
\end{align}
For photon energies up to 25 eV we used a radial grid from 0 to 140$a_0$ with ECS from 100$a_0$ ($a_0$ is the Bohr radius). 
At higher photon energies we used a smaller box of 70$a_0$ with ECS from 50$a_0$ and more dense grid. 
The effective one-particle Hamiltonian 
\begin{align}
\label{oneph}
h_{\ell}\left(r\right) = -\frac{\hbar^2}{2m}\frac{\partial^2}{\partial r^2} +
\frac{\hbar^2}{2m}\frac{\ell \left(\ell+1\right)}{r^2} 
\nonumber \\ -\frac{e^2}{4\pi \epsilon_0}\frac{Z}{r} + u_{{\rm HF}} + u_{{\rm proj}}
\end{align}
includes the (non-local) Hartree-Fock potential (HF), $u_{{\rm HF}}$,  for the closed shell with $N$ electrons and a correction, $u_{{\rm proj}}$ (also non-local), which ensures that any excited electron feels an approximate long-range potential with $N-1$ electrons remaining on the target. The latter potential is projected on virtual states and it therefore does not affect the occupied HF orbitals. Using the projected potential allows us to include some effects already in the basis set, that would otherwise be treated perturbatively through the RPAE-iterations. The eigenstates to $h_{\ell}$ form an orthonormal basis with eigen energies  $\epsilon_i$ that is used for the description of the occupied orbitals, but it is also used to span the virtual space of the photoelectron. The eigen energies of the virtual orbitals are complex in general due to ECS. 

As a first approximation an electron, originally in occupied state $a$ , that has absorbed ($+$), or  emitted ($-$) a photon of energy $\hbar \Omega$ can be described by a wave function
\begin{align}
\label{rho0}
\mid \rho^{(0)\pm}_{\Omega,a} \rangle  = 
\sum_{p}\frac{\mid p \rangle \langle  p \mid ez \mid a \rangle}{
\epsilon_a - \epsilon_p \pm  \hbar \Omega }E_{\pm\Omega}.
\end{align}
%where $d_{\Omega}$ denotes the dipole operator.
There is a pole in for the case of XUV absorption at $\epsilon_p = \epsilon_a + \hbar \Omega>0$.
An advantage with ECS is that the integration over the continuum is effectively performed along a different path in the complex energy plane and it is possible to replace the integration with a sum over a discretized representation of the excited states, $p$. The disadvantage of ECS is that the numerical representation of the perturbed wavefunction is physical only at radii before the complex scaling, $r<R_C$. 
%The 
%radial part of the 
%perturbed wave function $\rho^{(0)+}_{\Omega,a}$ in Eq.~(\ref{rho0})  is
%, after multiplication with the magnitude of the electric field, $E_{\Omega}$, 
%the first approximation to the radial wave function for the %outgoing electron.
% discussed in connection with Eq.~(\ref{onephoton}).

The many-body response to the interaction with the photon is neglected in Eq.~(\ref{rho0}), but the bulk of these effects can be added through the RPAE method, where certain sub-classes of many-body effects are included  through the iterative solution of the equations for the coupled channels:
\begin{align}
\label{RPA}
\mid \rho^{\pm}_{\Omega,a} \rangle
= \mid \rho^{(0)\pm}_{\Omega,a} \rangle 
-\sum_p^{exc} \frac{\mid p \rangle}{ \epsilon_a -\epsilon_p \pm \Omega} \nonumber \\
\times \Bigg(  \sum_b^{core}
\Bigg\{ 
\langle b p \mid V_{12} \mid a \,  \rho^{\pm}_{\Omega,b} \rangle
- \langle b \,  p \mid V_{12} \mid   \rho^{\pm}_{\Omega,b} \,  a \rangle \nonumber \\
+ \langle \rho^{\mp}_{\Omega,b} \,  p \mid V_{12} \mid    a b \rangle
- \langle p  \, \rho^{\mp}_{\Omega,b}  \mid V_{12} \mid    a b \rangle \Bigg\} \nonumber \\
-  \langle p  \mid u_{{\rm proj}}\left(r\right) \mid \rho^{\pm}_{\Omega,a} \rangle
\Bigg)
\end{align}
with the Coulomb interaction 
\begin{align}
V_{12} = \frac{e^2}{4\pi \epsilon_0}  \sum_{i <j}  \frac{1}{r_{ij}} = \frac{e^2}{4\pi \epsilon_0}  \sum_{i <j} 
\sum_K \frac{r_<^K}{r_>^{K+1}}  \mathbf{C}^K \left( i \right) \cdot \mathbf{C}^K\left( j \right).
\end{align}
Eq.~(\ref{RPA}) is illustrated by Goldstone diagrams in Fig.~\ref{fig:rpae}. The upper lines in Fig.~\ref{fig:rpae} show the diagrams for $\rho^{+}_{\Omega,a}$, where Fig.~\ref{fig:rpae}~(b) is the uncorrelated absorption of a photon $\Omega$ from Eq.~(\ref{rho0}). Fig.~\ref{fig:rpae}~(c) and (d) account for the electron--hole interaction in forward propagation,
while Fig.~\ref{fig:rpae}~(e) and (f) account for ground-state correlation effects. 
%
%and that in Fig.~\ref{fig:rpae}~c, with $K=0$  and $a=b$ ensures that the electron in $p$ sees the hole in $a$ 
%
When the projected potential is taken to be the monopole interaction with the outer-most hole $h$,  
\begin{align}
 u_{{\rm proj}}  =  \sum_{r,s}^\mathrm{virt.} \mid r \rangle \langle r h \mid \frac{1}{r_>} \mid s h \rangle \langle s  \mid,
\end{align}
%with $h$ being the hole left behind by the ionized or excited electron, 
the corresponding $K=0$ part of Fig.~\ref{RPA}~(c) are accounted for already in the basis sets and must be removed from the RPAE-expansion through the last term in Eq.~(\ref{RPA}) (not shown in Fig.~\ref{fig:rpae}). The RPAE iterations gives the same functions $\rho^{\pm}_{\Omega,a}$
regardless if the projected potential is used or not, but the convergence is often much improved in the latter case, especially close to the ionization threshold. 
%It is also possible to add part of the diagram in Fig.~\ref{fig:rpae}~d in the projected potential~\cite{lindroth:93:coll}, which sometimes  further improve the convergence. 
At sufficient distance from the core $\rho^{+}_{\Omega,a}$ will behave as an outgoing wave,
%although phase-shifted compared to that from a pure Coulomb potential,  
as indicated in Eq.~(\ref{onephoton2}), but in practice ECS leads to a damped wavefunction for $r > R_C$, as shown in  Fig.~\ref{fig:perturbedwf}~(a) and (b) for neon and fluorine, respectively. This damping occurs well outside the range of the bound electrons and both the photoionization cross section and the Wigner delay can be deduced from $\rho^{+}_{\Omega,a}$  in the unscaled region, $r<R_C$. 
%By checking the stability of the extracted numbers from several radial distances we can ensure that the numerical uncertainties are under control.

The affinity in the RPAE method is given by the HF orbital energies (following Koopmann's theorem), but it is customary to improve this description by instead using the experimental values. This approach is followed in the present study with substitutions according to Table~I. 
%where the F$^-$ HF energy of $4.9$~eV is replaced by $3.4$~eV and the Cl$^-$ HF energy of $4.1$~eV is replaced by $3.6$~eV. 

\begin{table}[!h]
\begin{center}
\begin{tabular}{c|cc}
 & HF  & EXP   \\ \hline 
F$^{-}$ & 4.9 & 3.4 \\ 
Cl$^{-}$& 4.1 & 3.6 \\
%\caption{(affinity in eV)}
\end{tabular}
\end{center}
\caption{Binding energy in eV.}
\end{table}

\subsection{Calculation of two-photon matrix elements}
When $\rho^{+}_{\Omega,a}$ is obtained an approximate two-photon matrix element for the first time-order (TO1) can be calculated as 
\begin{align}
\label{twophotonmatrix}
M^{(2,a/e)}_\textrm{[TO1]} = \langle f \mid ez \mid  \rho^{+}_{(2n\mp1)\omega,a} \rangle 
/E_{\Omega},
\end{align}
with a final state $f$ being an eigenstate to the effective one-particle Hamiltonian at the sideband kinetic energy $\epsilon = \epsilon_a + 2n\hbar\omega$. 
The numerical representation of radial part $u_{f,\ell}$ with angular momentum $\ell$ is then found by solution of  
\begin{align}
h_{\ell} u_{f,\ell}(r) = \epsilon u_{f,\ell}(r),
\end{align}
which can be reformulated as a system of linear equations  
for the unknown coefficients $c_i$ when expanded in B-splines
\begin{align}
 u_{f,\ell}(r) = \sum_i c_i B_i(r). 
\end{align}
The requirement that the solution is regular in the origin is enforced by exclusion of the only B-spline that is non-zero in the origin. The radial integration in Eq.~(\ref{twophotonmatrix}) is done numerically out to a distance far outside the atomic core within unscaled region ($a_0\ll r<R_C$), 
while the final part of the integral is carried out using analytical Coulomb waves % Since not only  $\rho^{+}_{\Omega,a}$, as discussed above,  but also $f$ will be a phase-shifted (Coulomb) waves here, the last part of the integral,  to $r \rightarrow \infty$,  can be calculated with analytical wave functions. This is done 
along the imaginary $r$-axis as described in Ref.~\cite{DahlstromJPB2014}. The numerical stability is monitored by comparison of  different ``break points'' between the numerical and analytical descriptions.

\begin{figure}
\includegraphics[width=0.47\textwidth]{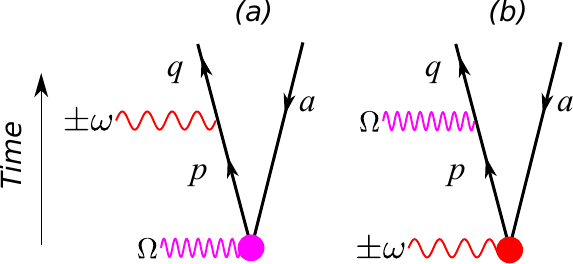}
\caption{
(a) First time-order process (TO1) where the XUV photon, $\Omega$, is absorbed before interaction with the laser field, $\pm\omega$. 
(b) Second time-order process (TO2) where the interaction with the laser occurs before absorption of the XUV photon.  
\label{fig:M2}}
\end{figure}
\subsection{The time-order of the photons}
The matrix element in Eq.~(\ref{twophotonmatrix}) is due to interaction with two photons but the result is not general because we assumed that the XUV photon absorption is followed by absorption or emission of a laser photon in an ATI-type process, shown in  Fig.~\ref{fig:M2}~(a). The total two-photon matrix element should be $M^{(2,a/e)}=M^{(2,a/e)}_\textrm{[TO1]}+M^{(2,a/e)}_\textrm{[TO2]}$ The other possibility, shown in Fig.~\ref{fig:M2}~(b), is that the laser photon is  absorbed or emitted first, creating initially a bound excited electron. Only after the second interaction will the electron be  ejected due to the large photon energy deposited by the XUV field. The two-photon matrix element for the second time-order (TO2) is 
\begin{align}
\label{twophotonmatrix2}
M^{(2,a/e)}_\textrm{[TO2]} = \langle f \mid ez \mid  \rho^{+}_{\pm\omega,a}
\rangle
/E_{\pm\omega}.
\end{align} 

\begin{figure*}
%\centering
\begin{minipage}{.5\textwidth}
  \includegraphics[width=.9\linewidth]{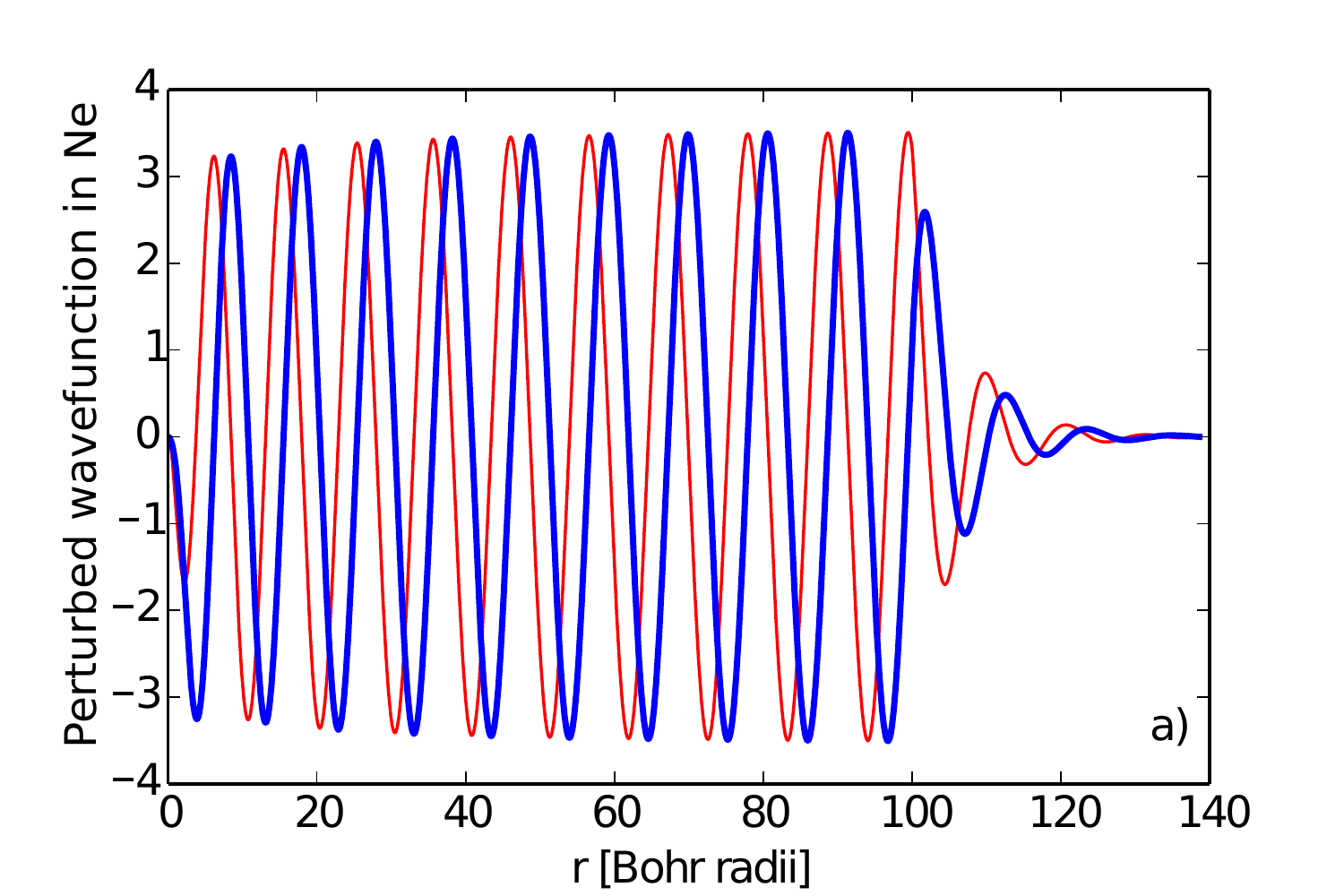}
  \includegraphics[width=.9\linewidth]{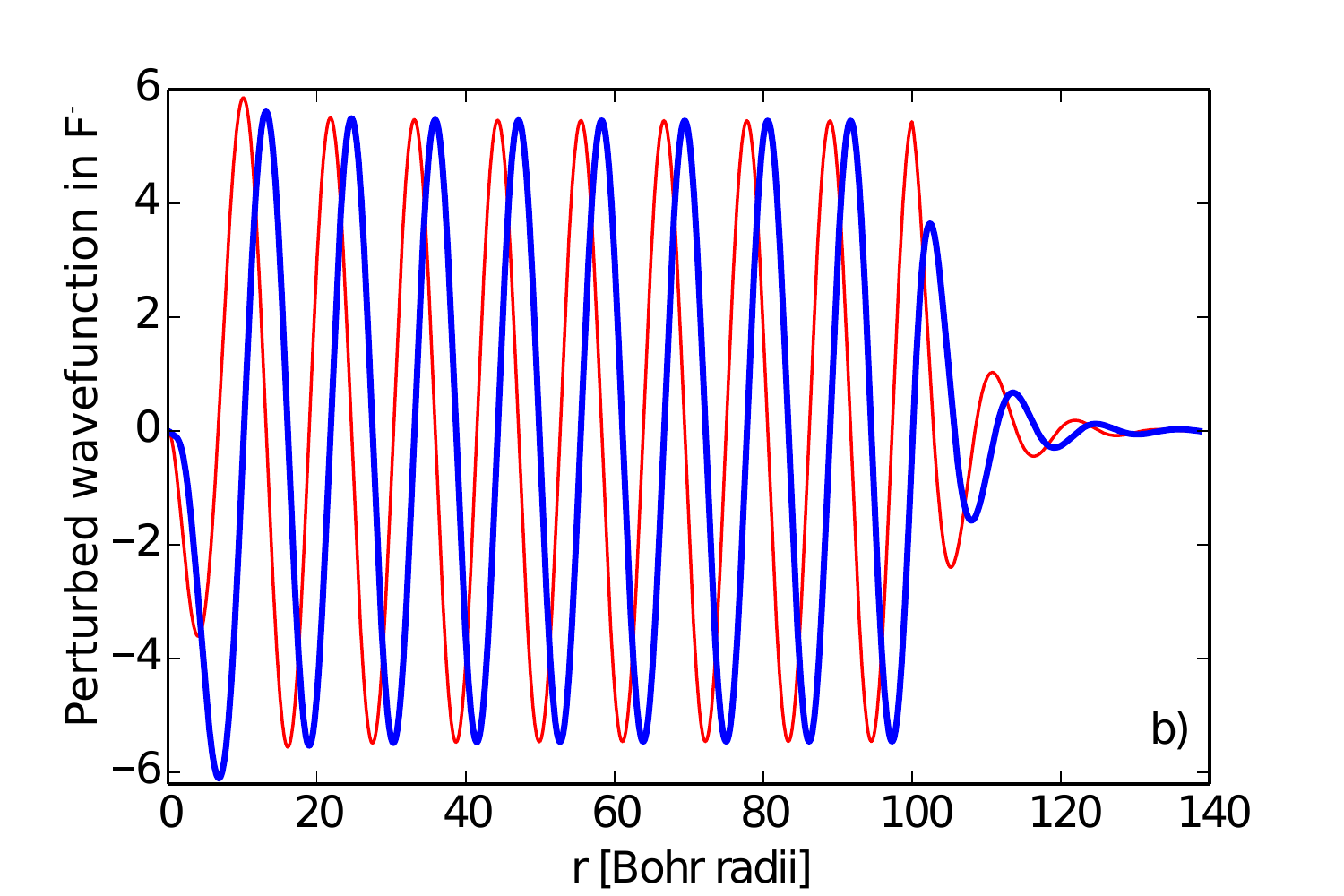}
\end{minipage}%
\begin{minipage}{.5\textwidth}
  \includegraphics[width=.9\linewidth]{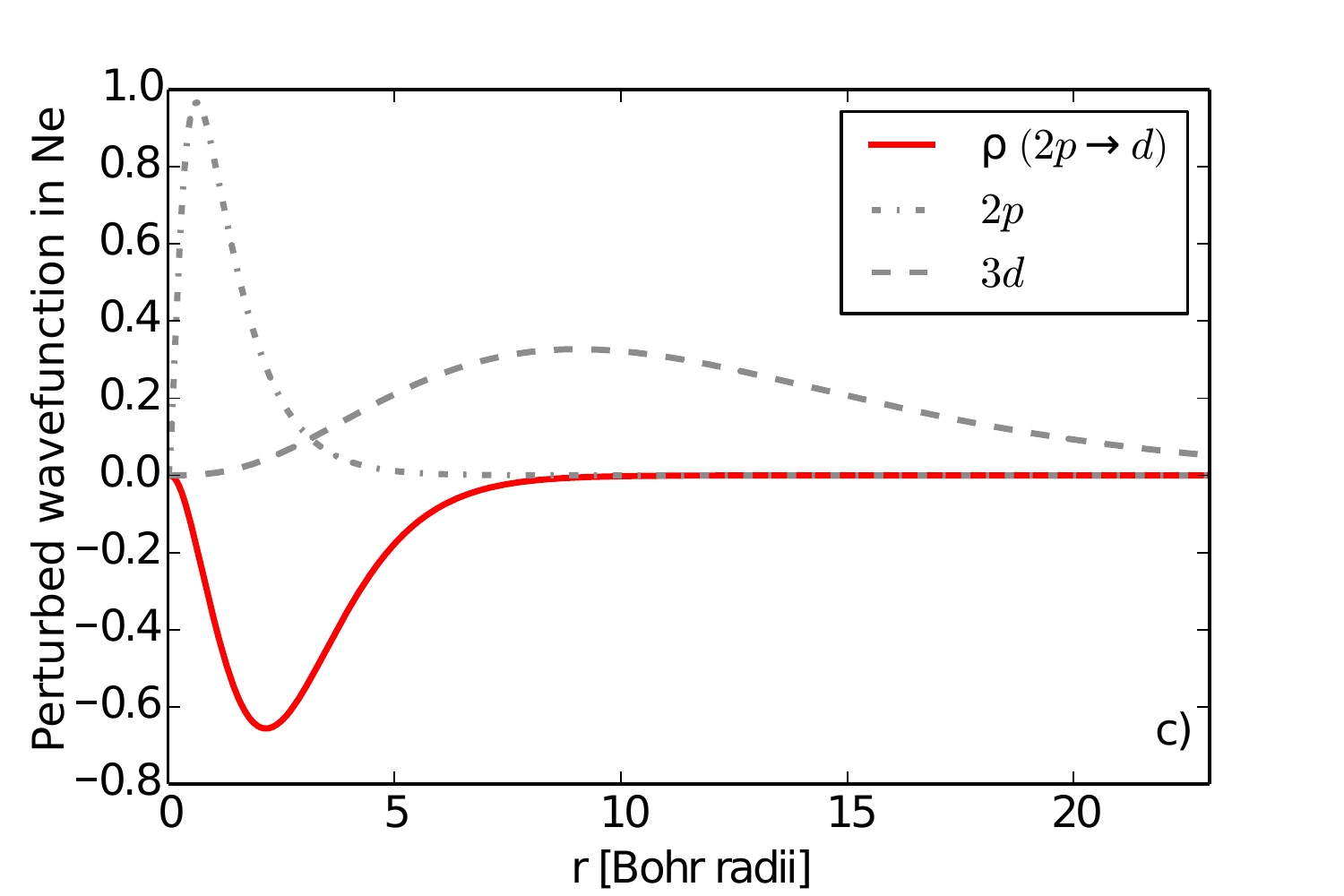}
    \includegraphics[width=.9\linewidth]{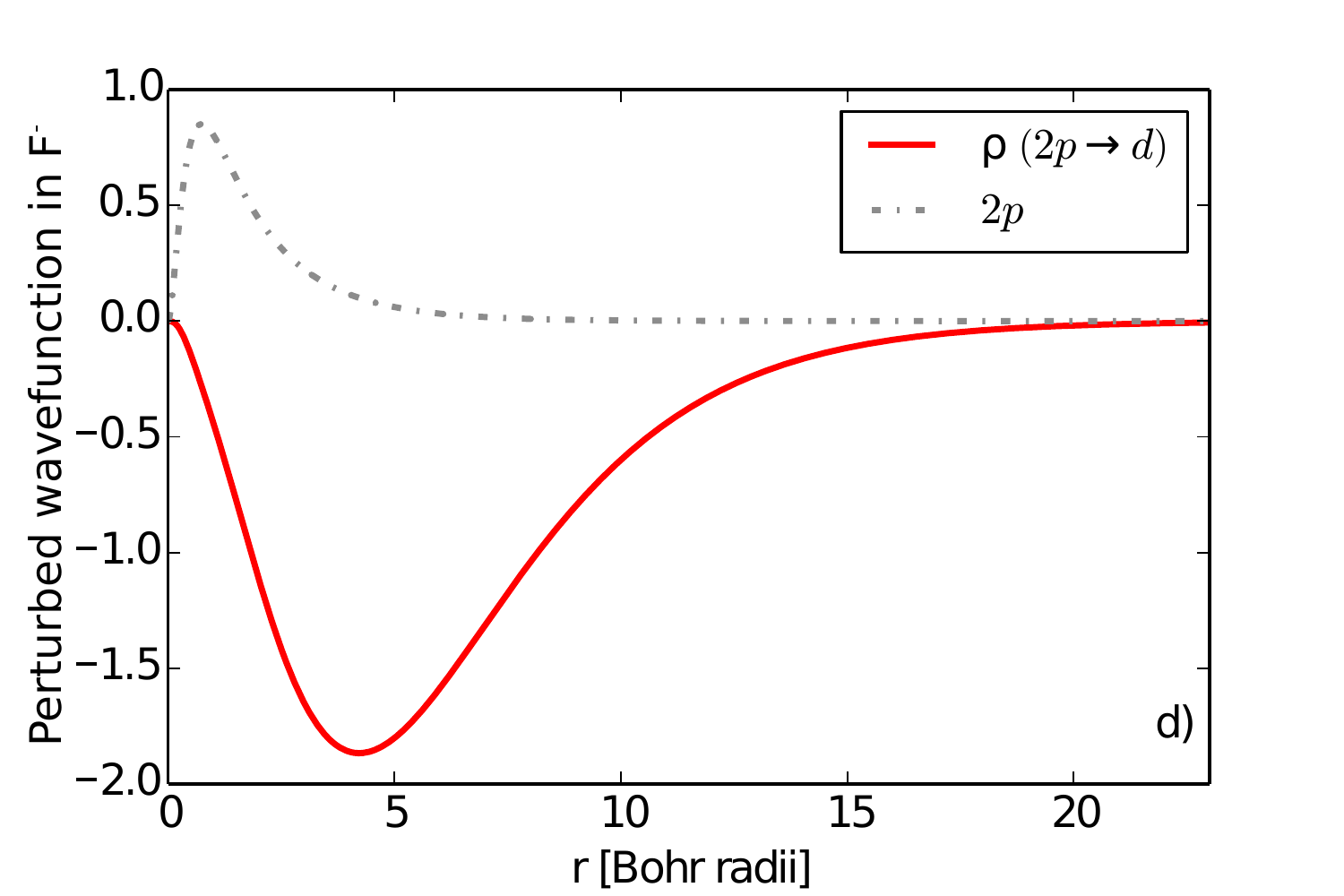}
\end{minipage}
\caption{The two left panels show the perturbed wave function $\rho \left( 2p \rightarrow d \right)$ for  Ne (a) and F$^-$ (b) after absorption of a photon with an energy resulting in a photoelectron of $4.3$~eV. The real part is shown with thin red lines and the imaginary part with thick blue lines.  The radial coordinate is complex rotated from $r=100$~a$_0$, which is the reason for the damping of the  perturbed wave function outside this distance. The two right panels show the perturbed wave function after absorption of a laser photon of $1.55$~eV (red solid line). The photon energy is below the ionization limit and results in a localized real correction to the wave function. The wave function for the initial state of the electron (the $2p$-orbital) is shown for comparison (dot-dashed gray line). In neon also the first excited state with $d$-character is shown.}
\label{fig:perturbedwf}
\end{figure*}

The very different nature of these two time-orders at the intermediate step is illustrated in Fig.~\ref{fig:perturbedwf}, where the left (right) panels show the perturbed wave function, $\rho \left( 2p \rightarrow d \right)$, for  Ne  and F$^-$  after absorption of an XUV (IR) photon. In the case of IR-first, we find that the wavefunctions extend further out than the initial occupied $2p$ orbitals, but not as far out as the virtual $3d$ orbital in neon (no such bound orbital exists in fluorine). The localization of the bound wavefunction is larger in neon than in fluorine. 
Differences between neon and fluorine are also visible in the case of XUV-first (in this example the electron is ejected to $4.3$~eV). The slow amplitude increases associated with Coulomb waves is observed in the case of neon [Fig.~\ref{fig:perturbedwf}~(a)], while it is absent in the case of fluorine, where instead a plane wave with uniform oscillations is established [Fig.~\ref{fig:perturbedwf}~(b)]. 

In general the second time-order is less important so that the total two-photon matrix element can be approximated by the first time-order. One reason for this is that there are no states that can be reached  by a resonant IR-transition from the ground state in either of studied systems. In the negative ions bound excited states are even completely absent. The contributing virtual transitions will then always be connected with large energy denominators, and this  will suppress the contribution. In previous studies on neon and argon~\cite{DahlstromPRA2012,DahlstromJPB2014} this time-order was in consequence with this neglected. Due to the smaller binding energies in the negative ions the process is less suppressed here. This is clearly seen in Fig.~\ref{fig:perturbedwf}~(d) where the correction in F$^-$ is significantly larger than that in neon in Fig.~\ref{fig:perturbedwf}~(c). 
%Also the contribution to the two-photon matrix element is much larger as will be in Sec.~\ref{results} below.

%\subsection{Gauge-invariance}
%The interaction with the electromagnetic field can be expressed in %different gauges: e.g. the length gauge %($e\mathbf{r}\cdot\mathbf{E}$) and the velocity gauge
%($(e/m)\mathbf{p}\cdot\mathbf{A}$), with $\mathbf{E}$ being the %electric field and $\mathbf{A}$ the vector potential.
%The RPAE-approximation is known to produce %gauge-%invariant~\cite{lin:77} results for the one-photon %transition %matrix elements. This holds when the approximation is %used without truncations (i.e. the sum over core orbitals  in %Eq.~(\ref{RPA})  should not be truncated).
%If the orbital energies are simply replaced with 
%experimental ionization energies this also destroys the 
%gauge invariance. This since the latter are not the true  %eigenvalues to the one-particle Hamiltonian.
%The gauge-issue in connection with  the atomic delay has not been %discussed earlier. Here we investigate the 
%result in the two gauges and find a very small gauge-dependence.
\begin{figure*}
%\centering
\begin{minipage}{.5\textwidth}
  \includegraphics[width=.9\linewidth]{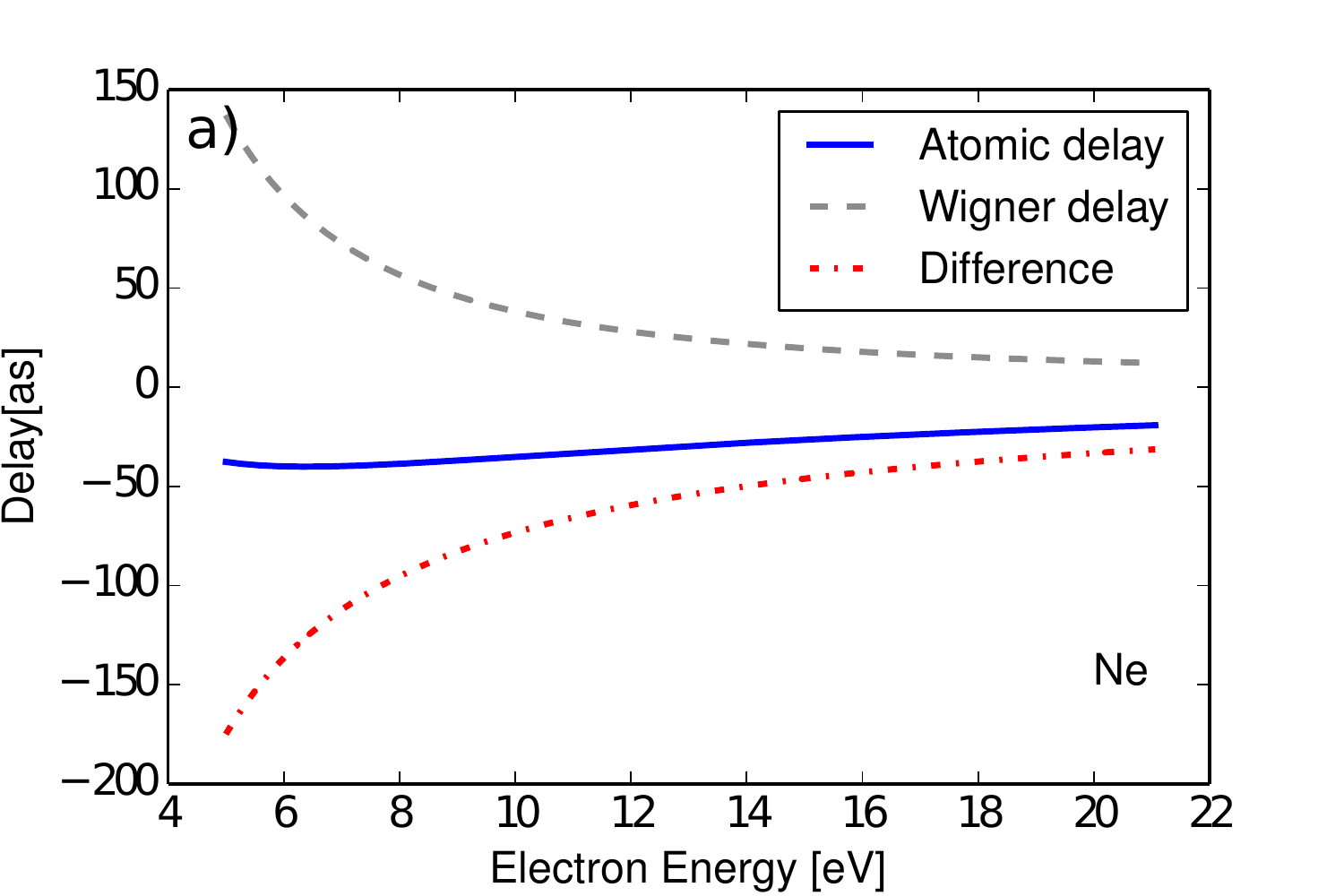}
  \includegraphics[width=.9\linewidth]{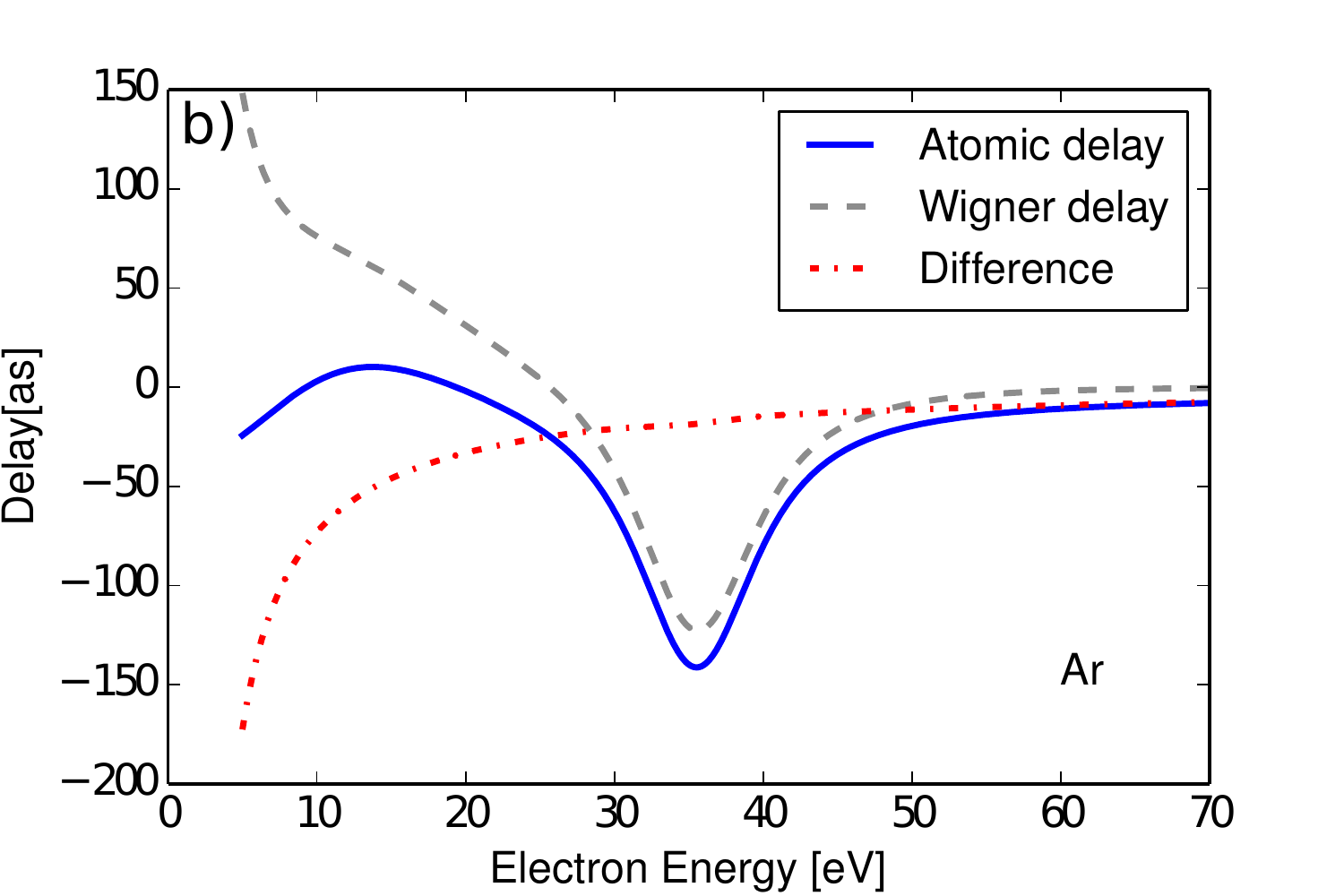}
\end{minipage}%
\begin{minipage}{.5\textwidth}
  \includegraphics[width=.9\linewidth]{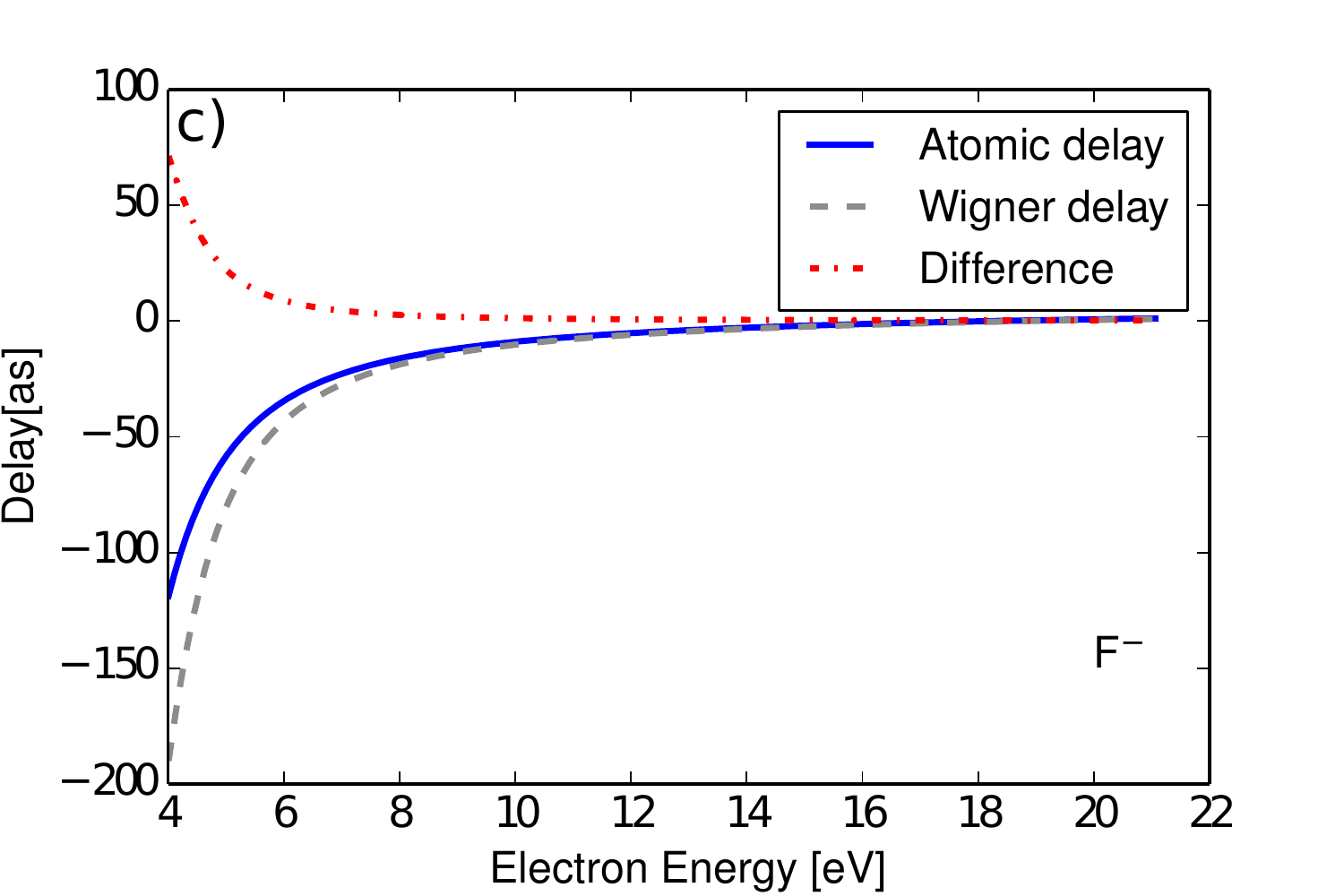}
    \includegraphics[width=.9\linewidth]{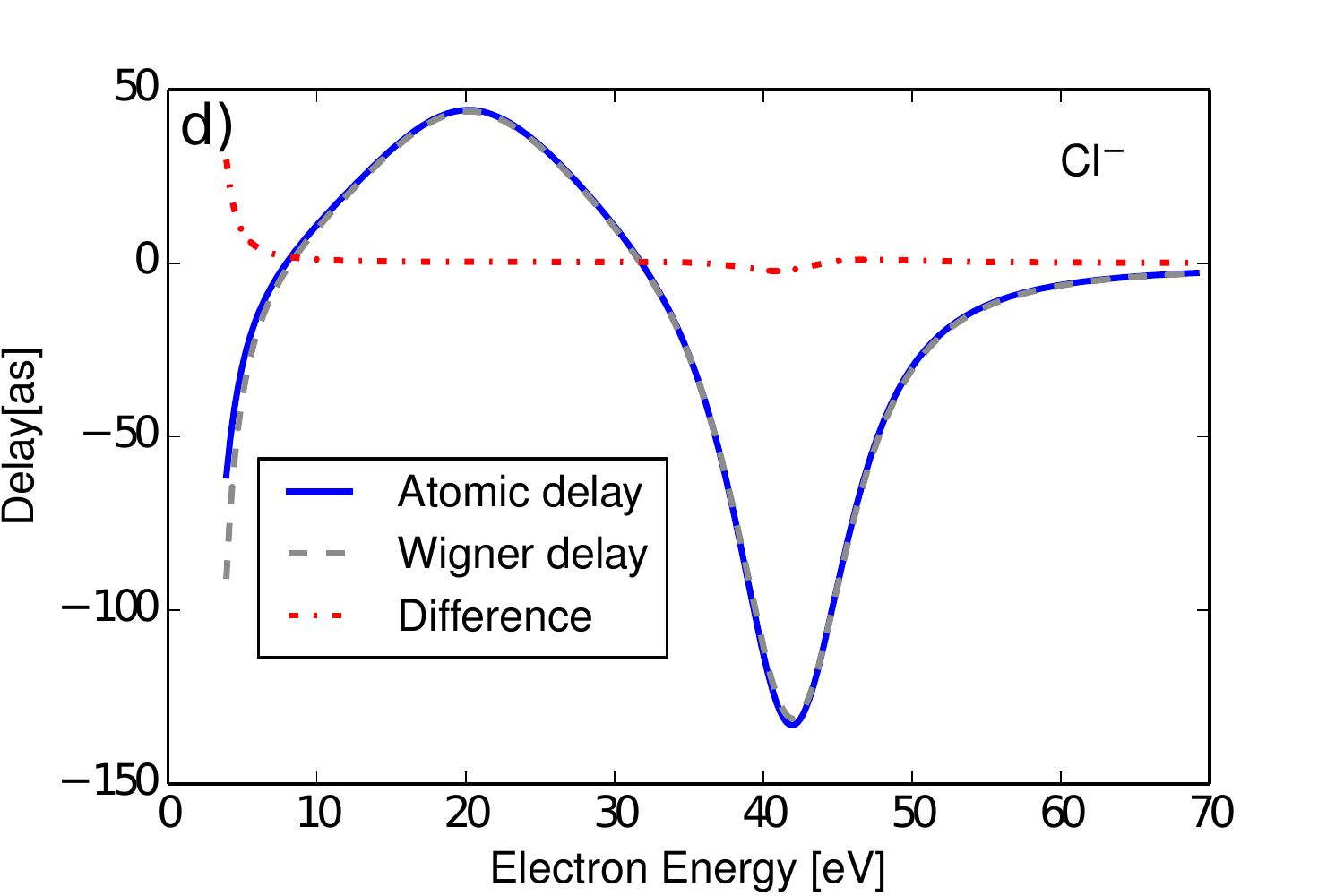}
\end{minipage}
\caption{The one-photon Wigner delay and the two-photon atomic delay for electron emission from the $2p$ -orbital in Ne and F$^-$, as well as from the $3p$ -orbital in Ar and Cl$^-$.}
\label{fig:delay}
\end{figure*}

\section{Results} 
\label{results}
The one-photon Wigner delay [Eq.~\ref{wignerdelay}] and the two-photon atomic delay [Eq.~(\ref{atomicdelay})] for electron emission along the common polarization axis $\theta=0$ are presented in Fig.~\ref{fig:delay} for neon, argon, fluorine chlorine including both time-orders (TO1+TO2). The fundamental laser photon energy $\hbar\omega$ is 1.55 eV. In the case of noble gas atoms, the delay difference -- known as the universal continuum--continuum delay, $\tau_{CC}$, induced by the second photon -- is substantial, negative and identical for neon and argon \cite{DahlstromPRA2012}. In the case of negative ions the delay difference is smaller and the Wigner delay is on top of the atomic delay at high kinetic energies. This is shown clearly for the case of Cl$^-$ where both delay curves exhibit the same negative delay peak of -140 as at the Cooper minimum close to 42 eV. At low kinetic energies, however, we find a clear delay deviation from zero that we will attribute to the short-range potential of the neutralized ion.  

\begin{figure}
\includegraphics[width=0.490\textwidth]{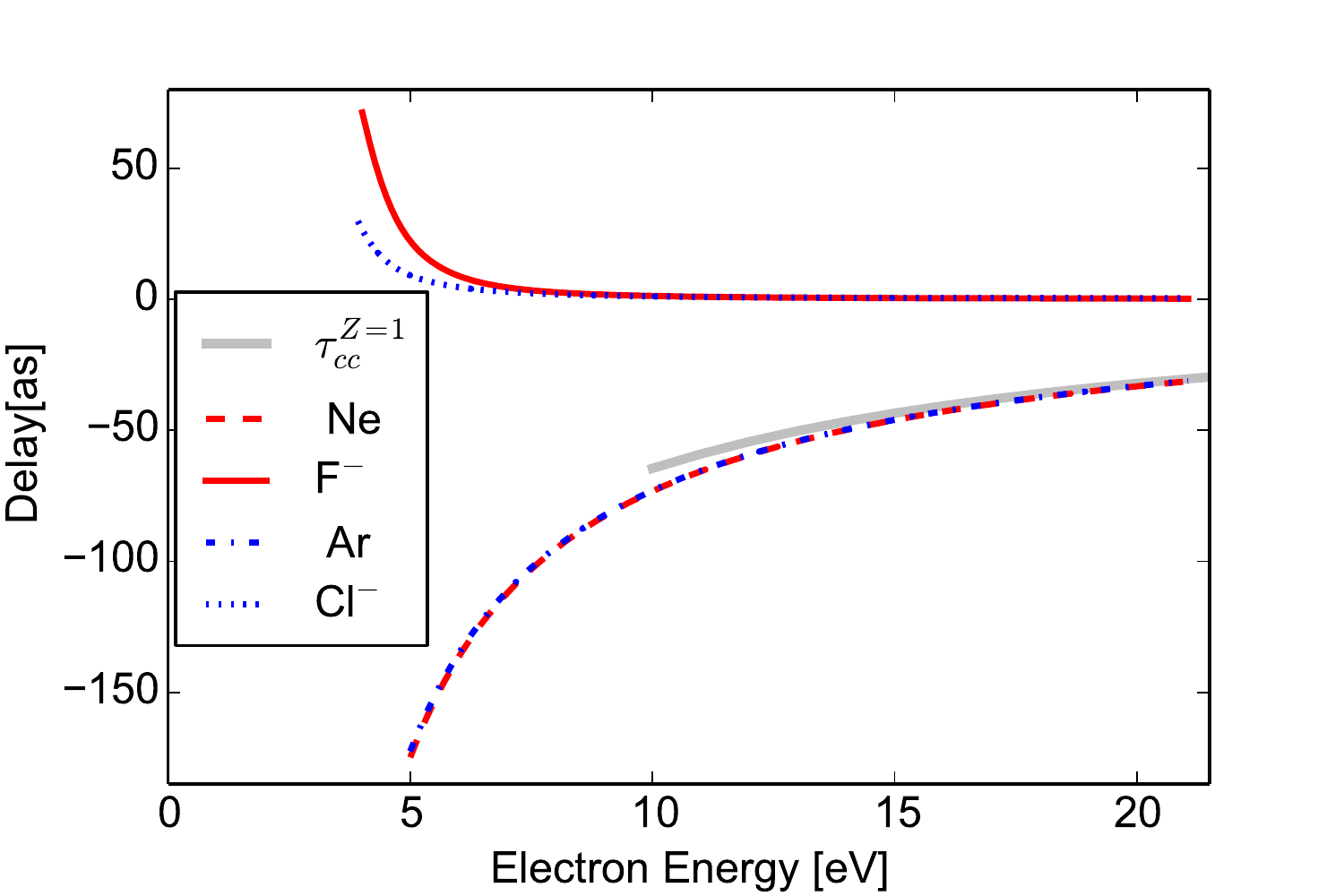}
\caption{
The difference between the two-photon atomic delay and the one-photon Wigner delay
for Ne, Ar, F$^-$ and Cl$^-$.\label{fig:compare}}
\end{figure}

For better comparison we show in Fig.~\ref{fig:compare} the delay differences and an estimated $\tau_{CC}^{Z=1}$ (obtained from the analytical expression Eq.~(100) in Ref.~\cite{DahlstromJPB2012} including both long-range phase and amplitude effects) for the case  Z=1. We stress that the agreement with the approximate $\tau_{CC}^{Z=1}$ is not perfect at low kinetic energies, but the agreement is good at high energies where the delay difference of the noble gas atoms is close to $\tau_{CC}^{Z=1}$. Similarly the prediction of the CC-theory is good for negative ions because the delay difference is close to zero (the estimated $\tau_{CC}^{Z=0}$ curve is not plotted since it is exactly zero).
%
%The time delay difference in F$^-$ shows a monotonic exponential decay at high kinetic energy. 
% $\sim \exp(-3.5 \epsilon[eV])$ (evaluated for the TO1-process).
% 
Interestingly, while the noble gas delay differences follow the same universal negative curve also at low energies, the negative ions depart from zero on different positive slopes. The departure is stronger in F$^-$ than in Cl$^-$ and it exceeds 50 as. This is an important result because it suggests that the universality found in noble gas atoms is not found in negative ions at low energies. 

\begin{figure*}
\begin{minipage}{.5\textwidth}
\includegraphics[width=0.90\linewidth]{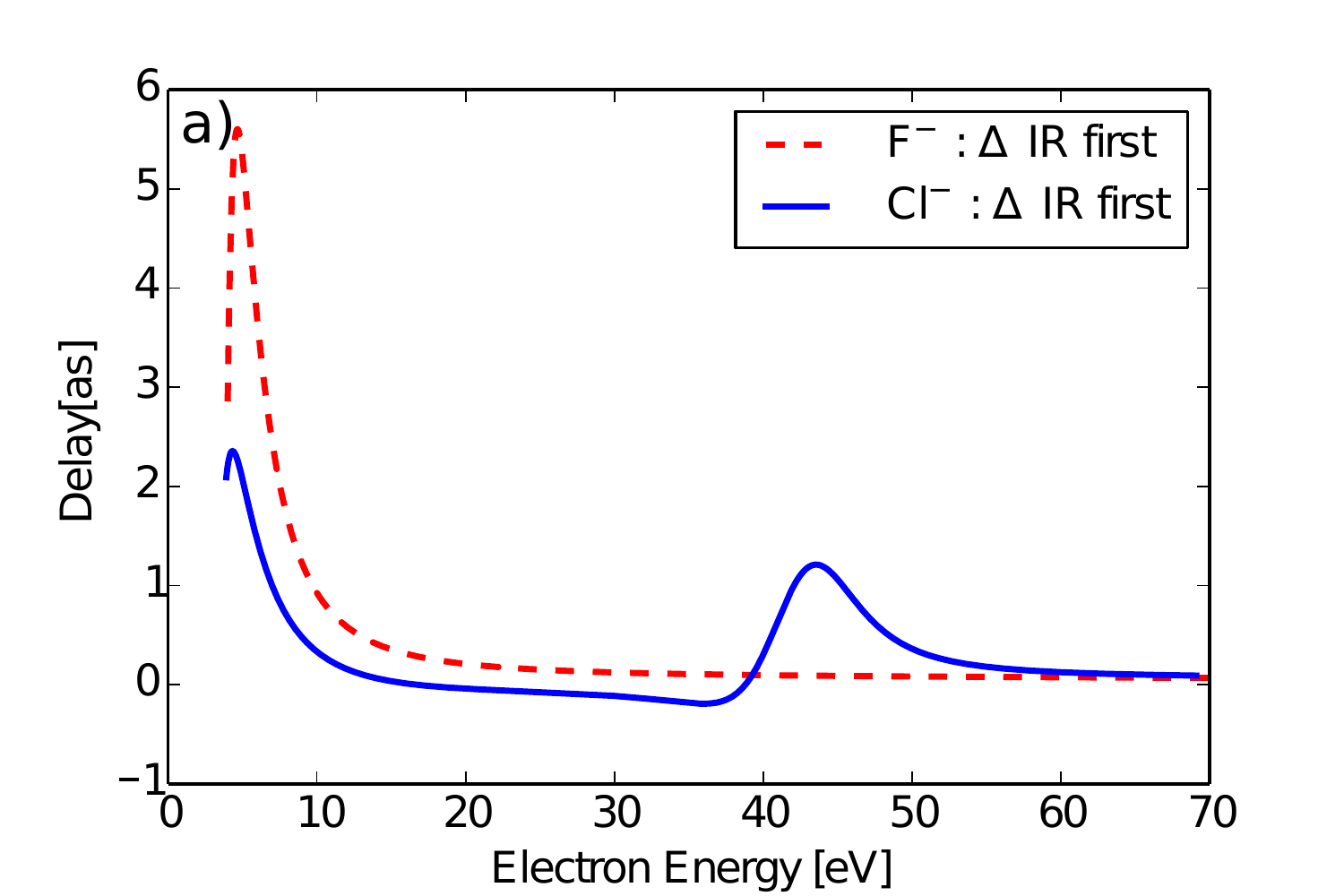}
\end{minipage}%
\begin{minipage}{.5\textwidth}
\includegraphics[width=0.90\linewidth]{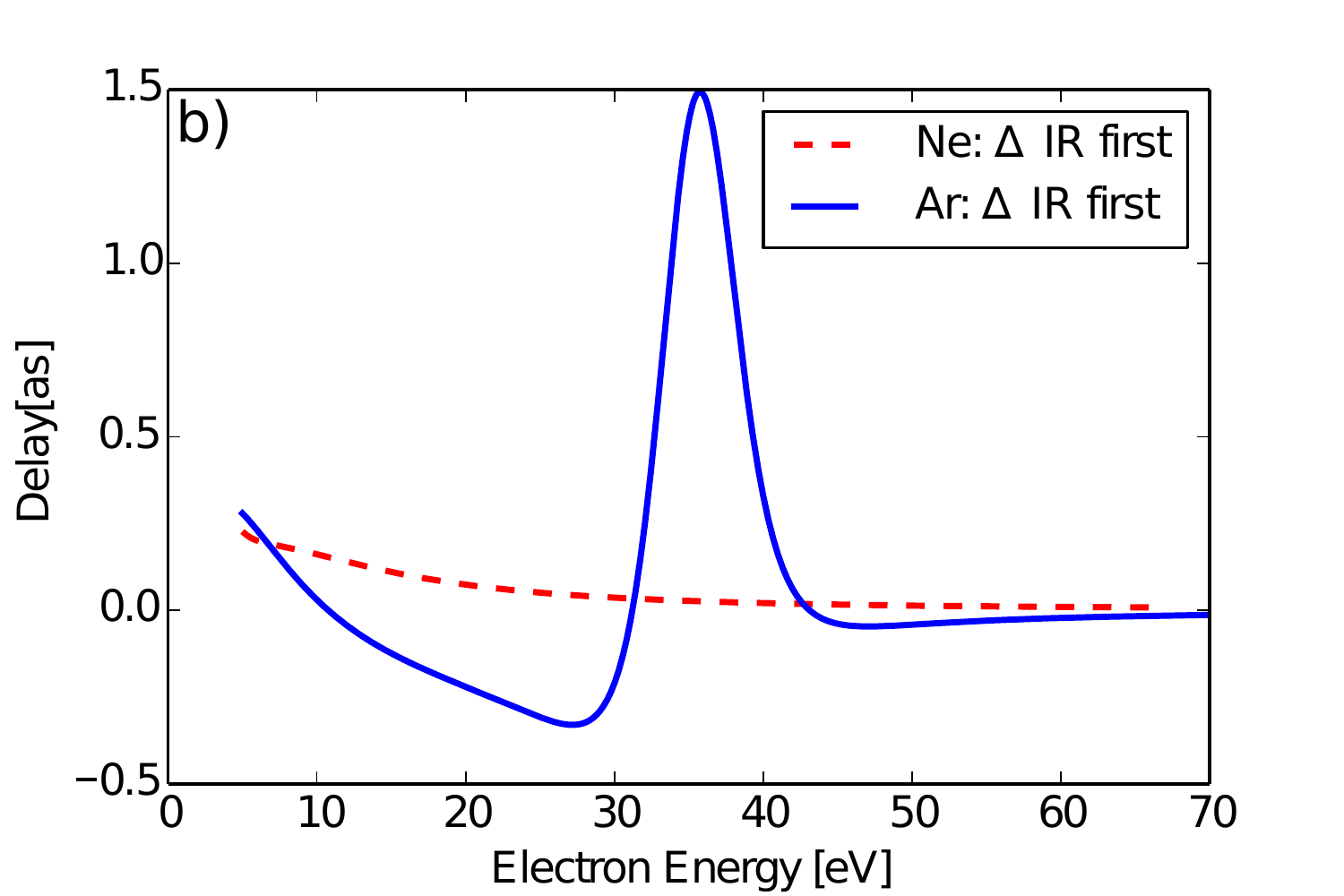}
\end{minipage}%
\caption{
The effect on $\tau_a - \tau_W$ from the second photon time-order. 
The plots show the difference between the difference delay with both time orders and that  with only the dominating time order (i.e. with the XUV photon being absorbed first)  
for negative ions F$^-$ and  Cl$^-$ (a) and atoms Ne and  Ar (b). \label{fig:timeorder} }
\end{figure*}

%\subsection{The time-order of the photons}
Next we show the difference of difference delay between all time-orders (TO1+TO2) and the dominant time-order (TO1) in Fig.~\ref{fig:timeorder}. As expected the effect of the second time-order in noble gas atoms is small and it amounts to less than 1.5 attoseconds in argon and even less in neon. At high energies the time-order difference on an absolute scale are similar in each isoelectronic negative ion. e.g. the same peak height of 1.5 attoseconds is observed in Cl$^-$. This effect can be attributed to the Cooper minimum around 42 eV (35 eV) that occurs in the $p \rightarrow d$ ionization step in chlorine (argon). At this energy the otherwise subordinate $p \rightarrow s$ ionization channel will dominate and it will interfere the even weaker TO2-processes. On the other hand, because the difference between atomic delay and Wigner delay is much smaller in the negative ions the relative importance of the TO2-processes is larger than in neutral atoms; around 10 percent instead of a few per mille. 

\begin{figure}
\includegraphics[width=0.490\textwidth]{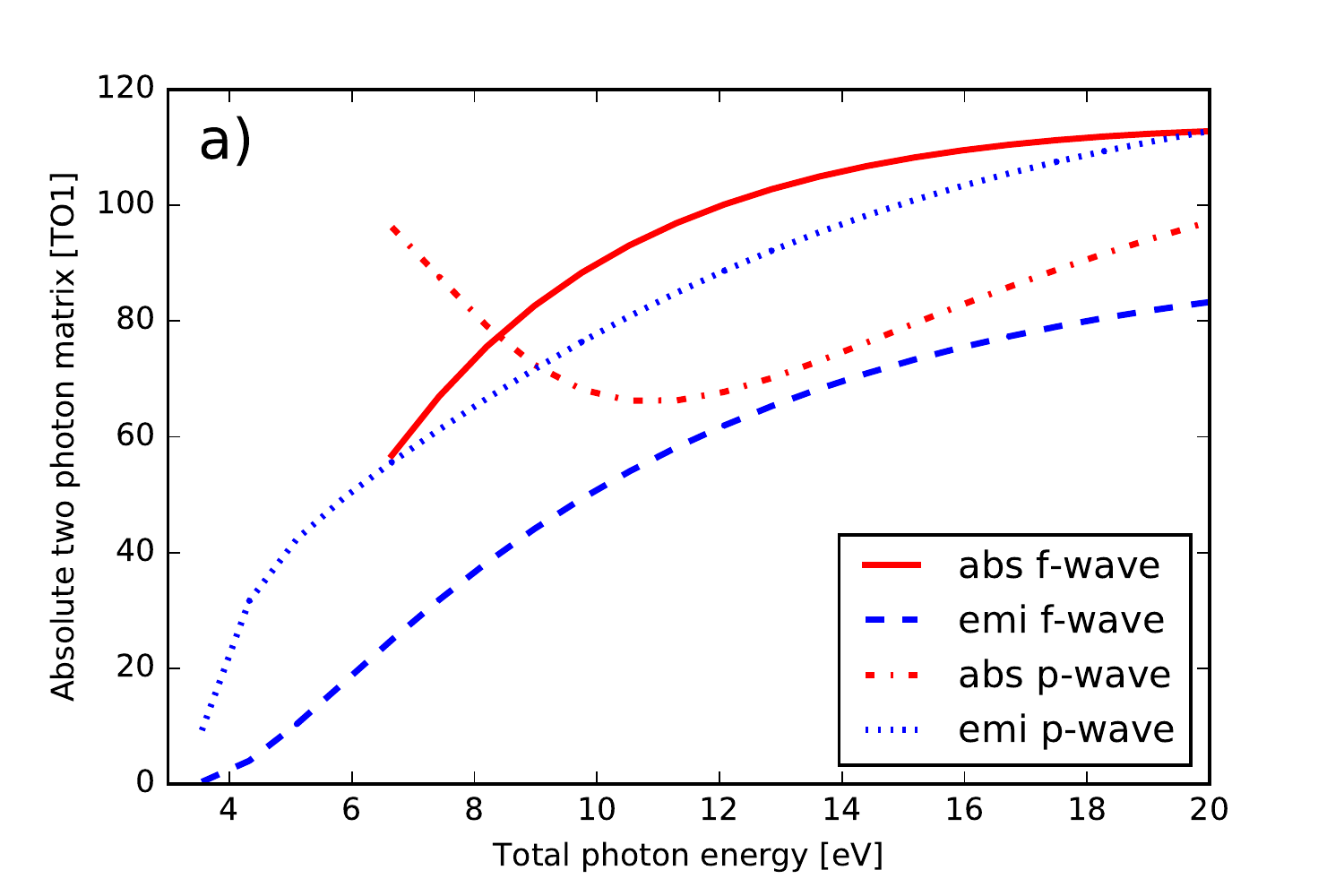}
\includegraphics[width=0.490\textwidth]{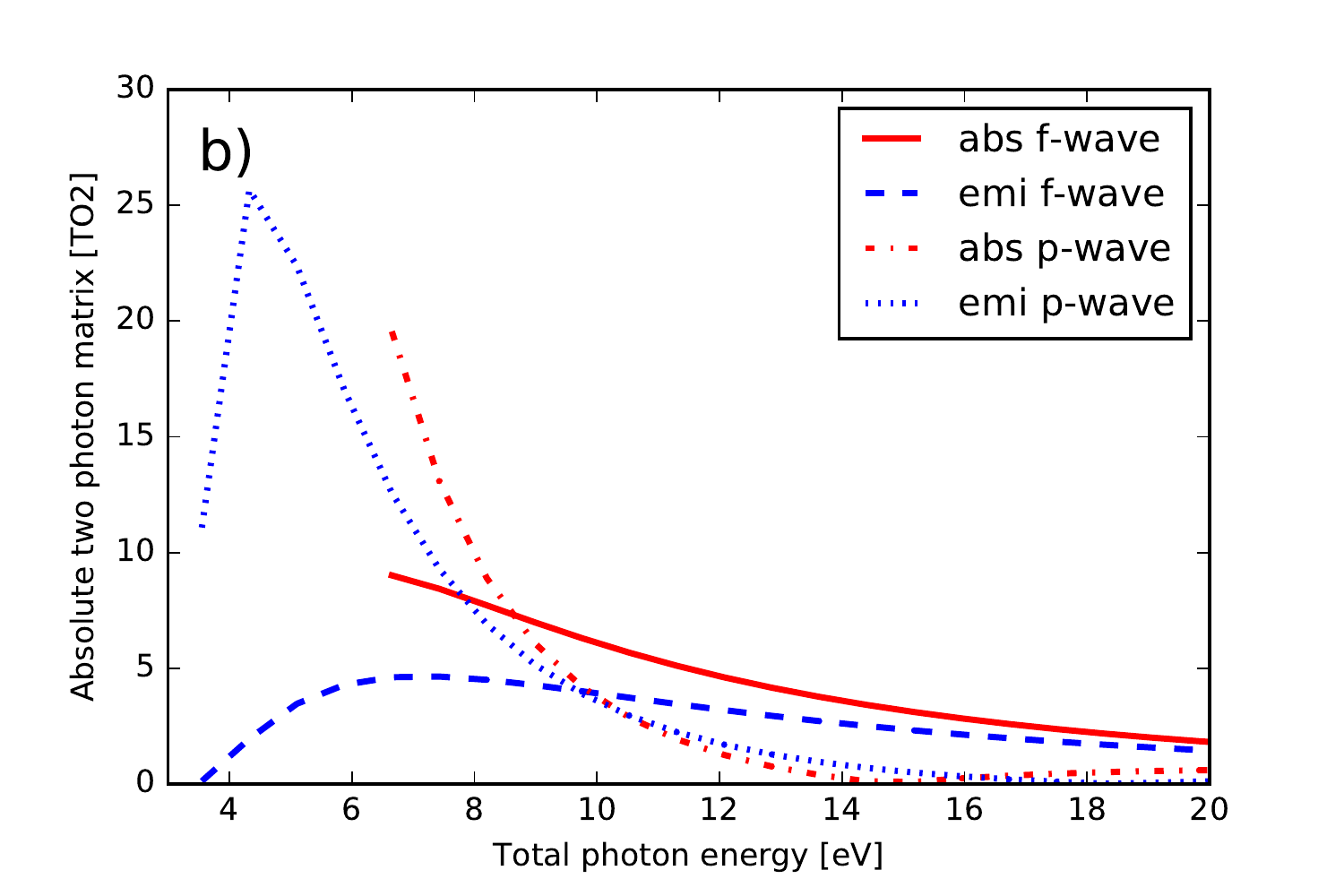}
\caption{(a) Absolute value of two-photon matrix elements with XUV photon absorbed before the laser interaction (process denoted TO1 in main text). The contributions on final $p$ and $f$-waves are shown separately with both absorption and emission of laser photon. (b) Same as (a) except that the XUV photon is absorbed after the laser interaction (TO2). }
\label{fig:absM2}
\end{figure}

At low kinetic energies we find that the negative ions exhibit a stronger dependence on the TO2-processes (up to a few attoseconds) also on the absolute time scale. The larger delay effects due to TO2 is expected at the threshold because (i) the affinities are comparable to the IR laser photon energy and (ii) the one-photon cross section vanishes at the ionization threshold of negative ions. Given (ii) one could expect that TO1 should be strongly reduced because it is initiated by absorption of an XUV photon. However, the two-photon amplitude effects are rather subtle as shown in Fig.~\ref{fig:absM2}. Both absolute values of TO1 and TO2 processes with emission of a laser photon (blue curves) approach zero as the total photon energy approaches the affinity threshold (3.4 eV) in Fig.~\ref{fig:absM2}~(a) and (b), respectively, where the absorbed XUV photon energy starts at 5.1~eV and the emitted laser photon is 1.55~eV. In the energy range shown, the TO1 emission processes monotonically approach zero, while the TO2 processes undergo a peak before going to zero at the threshold. The peak is especially strong for the $p$-wave and a crossing of the absolute values of $p$ and $f$-waves occurs close to 10~eV. Considering the same XUV frequencies, the corresponding TO1 and TO2 processes with absorption of a laser photon (red curves) lead to larger photoelectron energies and to non-zero absolute values. However, we have checked that the TO2 absorption curves also approach zero when the total photon energy approaches the affinity energy. Clearly the TO2 processes can be comparable to the TO1 processes close to the threshold, but it does not account for the 10s of as observed in Fig.~\ref{fig:compare}~(c). 

Second, we point out a peculiar behavior of the TO1 laser absorption process with a low energy $p$-wave in Fig~\ref{fig:absM2}. We observe a sudden increase in $p$-wave amplitude and a crossing with the f-wave close to 9~eV as the total photon energy is decreased. We propose that the effect arises due to the $p-s-p$ pathway that allows the electron to go close to the ionic core to absorb the laser photon, while the $p-d-p$ and $p-d-f$ are further out, due to the positive centrifugal barrier, and therefore not able to absorb a laser photon as efficiently. The same effect is not observed in the isoelectronic neon system, where the long-range Coulomb potential allows for laser-driven transitions further out from the atomic core. We propose that this strong dependence on the exact short-range potential explains the breakdown of the universality of the delay differences shown in Fig.~\ref{fig:compare} for negative ions.      

\section{Conclusion}
\label{sec:conclusions}
We have computed laser-assisted photoionization time delays from closed-shell negative ions with the aim to quantify the so-called continuum-continuum delay, $\tau_{CC}$, in finite systems that lack long-range interaction. In agreement with earlier analytical predictions, based on the asymptotic form of the wave functions \cite{DahlstromCP2013}, we find that the CC-delay from negative ions is negligible at high photoelectron energy when compared to the one-photon Wigner-like delay. While the CC-delay is not exactly zero, it is found to decrease exponentially -- much faster than in atoms -- so that the Wigner delay can be directly accessible in experiments. At low photoelectron energies, where the asymptotic form is expected to break down, we find that the CC-delay increases sharply in both studied negative ions. Further, we find that there is no universal CC-delay for F$^-$ and Cl$^-$ ions, despite the fact that a universal CC-delay has been reported in the isoelectronic Ne and Ar atoms \cite{DahlstromPRA2012} at the same low kinetic energy range. We attribute this strong CC-delay close to the threshold to the fact that absorption of laser photons at low kinetic energies is greatly increased for the final $p$-wave state that involves a virtual $s$-wave close the core. In practice, because the coupling between laser and short-range potential is system specific, a simple extraction of the Wigner delay is not possible experimentally. Finally, we have estimated the effects of reversed time-order processes, where IR photons are exchanged before the absorption of the XUV harmonics. While these effects also increase at low photoelectron energy, they remain smaller than the non-universal CC-delay in both studied negative ions. 
%We have found that the laser-induced delay in RABBITT type experiments -- referred to as the continuum-continuum delay $\tau_{CC}$ -- is not infinitely small or universal when an electron leaves a neutralized target with low kinetic energy. We report a significant delay for the case of low energy photodetachment of the negative ions fluorine and chlorine and show that absorption of laser photons at slow kinetic energies is greatly increased for the final $p$-wave state. At higher electron energies we find that the laser-induced delay is decreases  exponentially -- much faster than in atoms -- so that the one-photon Wigner delay should can be directly experimentally accessible. 
\label{Conclusion}
\section*{Acknowledgments}
EL acknowledges support from the Swedish Research
Council, Grant No. 2016-03789 and JMD  from 
Swedish Research Council, Grant No. 2014-3724.
%
%\bibliography{cc,references}

\end{document}